\documentclass{article}
\usepackage{authblk}
\usepackage[utf8]{inputenc}
\usepackage{geometry}
\usepackage{bm}
\usepackage{rotating}
\usepackage{caption}
\usepackage{subcaption}
\usepackage{amsmath,amssymb,mathtools,amsthm,longtable}
\usepackage{natbib}
\usepackage{cases}
\usepackage[hyphens]{url}
\usepackage[linesnumbered, ruled]{algorithm2e}
\usepackage{hyperref}
\hypersetup{
colorlinks=true,
linkcolor=red,
filecolor=magenta,
citecolor=blue,
urlcolor=blue,
}
\usepackage{cleveref}
\Crefname{assumption}{Assumption}{Assumptions}
\usepackage{setspace}
\usepackage{appendix}

\usepackage{bbm} 
\usepackage{textcomp} 

\usepackage{multirow,multicol,makecell,booktabs}

\usepackage{xargs} 
\usepackage[pdftex,dvipsnames]{xcolor}
\usepackage[colorinlistoftodos,prependcaption,textsize=tiny]{todonotes}
\newcommandx{\change}[2][1=]{\todo[linecolor=blue,backgroundcolor=blue!25,bordercolor=blue,#1]{#2}}
\newcommandx{\info}[2][1=]{\todo[linecolor=OliveGreen,backgroundcolor=OliveGreen!25,bordercolor=OliveGreen,#1]{#2}}
\newcommandx{\improvement}[2][1=]{\todo[linecolor=Plum,backgroundcolor=Plum!25,bordercolor=Plum,#1]{#2}}
\newcommandx{\thiswillnotshow}[2][1=]{\todo[disable,#1]{#2}}

\newcommand{\ubar}[1]{\mkern 1.5mu\underline{\mkern-1.5mu#1\mkern-1.5mu}\mkern 1.5mu}

\def\bgamma{\bm{\gamma}}

\def\btheta{\bm{\theta}}

\def\Exp{\mathrm{E}}
\def\Var{\mathrm{Var}}

\def\diag{\mathrm{diag}}

\def\rd{\mathrm{d}}

\def\bZ{\mathbf{Z}}
\def\ba{\bm{a}}
\def\bb{\mathbf{b}}

\def\be{\mathbf{e}}
\def\bg{\mathbf{g}}

\def\br{\mathbf{r}}

\def\bv{\mathbf{v}}
\def\bw{\mathbf{w}}

\def\cD{\mathcal{D}}

\def\cN{\mathcal{N}}

\def\cT{\mathcal{T}}

\def\cV{\mathcal{V}}

\def\diag{\mathrm{diag}}
\def\expit{\mathrm{expit}}

\title{Leveraging Neural Networks to Profile Health Care Providers with Application to Medicare Claims}
\author[1,2,3]{Wenbo Wu}
\author[4,5]{Fan Li}
\author[1]{Richard Liu}
\author[6]{Yiting Li}
\author[6,7]{Mara McAdams-DeMarco}
\author[8,3,9]{Krzysztof J. Geras}
\author[10]{Douglas E. Schaubel}
\author[1]{Iv\'an D\'iaz}
\affil[1]{Division of Biostatistics, Department of Population Health, NYU Grossman School of Medicine}
\affil[2]{Division of Nephrology, Department of Medicine, NYU Grossman School of Medicine}
\affil[3]{Center for Data Science, New York University}
\affil[4]{Department of Biostatistics, Yale School of Public Health}
\affil[5]{Center for Methods in Implementation and Prevention Science, Yale School of Public Health}
\affil[6]{Department of Surgery, NYU Grossman School of Medicine}
\affil[7]{Division of Epidemiology, Department of Population Health, NYU Grossman School of Medicine}
\affil[8]{Department of Radiology, NYU Grossman School of Medicine}
\affil[9]{Department of Computer Science, Courant Institute of Mathematical Sciences, New York University}
\affil[10]{Department of Biostatistics, Epidemiology and Informatics, University of Pennsylvania Perelman School of Medicine}
\date{\today}

\begin{document}

\maketitle

\begin{abstract}
Encompassing numerous nationwide, statewide, and institutional initiatives in the United States, provider profiling has evolved into a major health care undertaking with ubiquitous applications, profound implications, and high-stakes consequences. In line with such a significant profile, the literature has accumulated a number of developments dedicated to enhancing the statistical paradigm of provider profiling. Tackling wide-ranging profiling issues, these methods typically adjust for risk factors using linear predictors. While this approach is simple, it can be too restrictive to characterize complex and dynamic factor-outcome associations in certain contexts. One such example arises from evaluating dialysis facilities treating Medicare beneficiaries with end-stage renal disease. It is of primary interest to consider how the coronavirus disease (COVID-19) affected 30-day unplanned readmissions in 2020. The impact of COVID-19 on the risk of readmission varied dramatically across pandemic phases. To efficiently capture the variation while profiling facilities, we develop a generalized partially linear model (GPLM) that incorporates a neural network. Considering provider-level clustering, we implement the GPLM as a stratified sampling-based stochastic optimization algorithm that features accelerated convergence. Furthermore, an exact test is designed to identify under- and over-performing facilities, with an accompanying funnel plot to visualize profiles. The advantages of the proposed methods are demonstrated through simulation experiments and profiling dialysis facilities using 2020 Medicare claims from the United States Renal Data System.
\end{abstract}

\noindent%
\vfill

\newpage

\section{Introduction}

Health care provider profiling is a care quality assessment process routinely carried out by health care administrators and regulatory agencies \citep{welch1994physician,auerbach1999principles}. Throughout the process, the performance of clinicians, hospitals, or other types of providers is quantified and compared through standardized quality measures based on a variety of patient-centered outcomes, such as 30-day hospital readmission and death after hospital discharge. Outlying providers with significantly subpar services in terms of quality metrics are then identified, leading to increased public awareness, enhanced evidence-based accountability, and more targeted interventions for quality improvement. In the United States, provider profiling has been recognized as a useful tool for evaluating health care practitioners and institutions to promote coordinated and cost-effective quality care \citep{goldfield2003profiling}. As one of the first profiling programs, the New York State Department of Health stands as a pioneer in evaluating hospitals statewide conducting coronary artery bypass graft (CABG) surgeries and percutaneous coronary interventions since 1989 and 1996, respectively \citep{racz2010bayesian}. Nationally, the Medicare Prescription Drug, Improvement, and Modernization Act established the Hospital Inpatient Quality Improvement Program in 2003, urging hospitals to report 30-day all-cause readmission and mortality rates for acute myocardial infarction (AMI), heart failure (HF), and pneumonia; the Affordable Care Act launched the Hospital Readmissions Reduction Program in 2012, financially penalizing hospitals for excess readmissions in conditions like AMI, CABG surgery, and HF \citep{cmshiqip, cmshrrp}. In addition, the U.S. Centers for Medicare and Medicaid Services (CMS) administers the end-stage renal disease (ESRD) Quality Incentive Program to evaluate Medicare-certified kidney dialysis facilities providing services to Medicare beneficiaries with ESRD who require dialysis to survive; a facility with unsatisfactory performance (e.g., whose patients have experienced a readmission rate much higher than expected) will receive substantial payment reduction as a penalty \citep{cmsesrdqip}.

The widespread applications, far-reaching implications, and high stakes underscore the necessity for principled statistical and data science methods to improve the practical landscape of provider profiling, as noted in the pivotal white paper commissioned by the Committee of Presidents of Statistical Societies and CMS \citep{ash2012statistical}. Thus far, the literature has accumulated a burgeoning body of research aimed at advancing the methodology of provider profiling. Diverse statistical techniques have been employed to analyze longitudinal and time-to-event data in various profiling contexts, including the generalized linear (mixed) models \citep{normand1997statistical, ohlssen2007hierarchical, racz2010bayesian, ash2012statistical, he2013evaluating, kalbfleisch2013monitoring, estes2020profiling, xia2022accounting, wu2022improving, wu2023test}, (semi-)competing risk models \citep{lee2016hierarchical, wu2022analysis, lee2022facility, haneuse2022measuring}, inverse probability weighting \citep{tang2020constructing}, and a Bayesian finite mixture of global location models \citep{silva2023reformulating}. The vast majority of existing profiling methods have their roots in the fixed- and random-effects frameworks, in which the inter-provider variation of care quality is captured by either fixed or random effects, adjusting for patient characteristics and other relevant confounders. Despite the long-standing debate regarding their respective strengths and weaknesses in estimation \citep{he2013evaluating, kalbfleisch2013monitoring, kalbfleisch2018discussion}, both approaches assume that the effects of risk factors are linear, which can be too restrictive to characterize potentially dynamic effect trajectories or complex nonlinear relationships in practice. For instance, the coronavirus disease 2019 (COVID-19) pandemic has compelled CMS to adjust for COVID-19 in the development and maintenance of standardized quality measures. Tasked by CMS, our recent investigation suggests that the impact of COVID-19 on Medicare dialysis patients has dramatically evolved since the onset of the pandemic \citep{wu2022impact, wu2022understanding}. To this end, flexible risk adjustment models that relax the linearity assumption have great potential to better capture the nuanced effect variation and to improve the practice of providing profiling.

Deep learning, featuring flexible compositions of numerous nonlinear functions, has emerged as a leading tool that accommodates complex input-output relationships. It has embraced tremendous success in statistics, data science, computational medicine, health care, and other disciplines \citep{fan2021selective}. This success suggests that neural networks hold strong potential for surpassing the limitations of current profiling methods. However, several challenges arise when incorporating a neural network architecture into a profiling context. Firstly, many methods harnessing neural networks require that subject-level observations be independent \citep{mandel2023neural}. These deep learning methods cannot be immediately employed in profiling since subjects are naturally clustered by providers. Secondly, within the realm of deep learning methods that do not rely on the independence assumption, certain approaches solely target continuous outcomes \citep{tandon2006neural}, while others are primarily designed for prediction and classification \citep{tran2017random, mandel2023neural, simchoni2023integrating}, thus not directly applicable to profile providers.
As a last technical note, incorporating a large number of provider-specific effects in a neural network model using conventional optimization methods such as the stochastic gradient descent (SGD) can lead to prolonged time to convergence. These methods typically rely on simple random sampling to update model parameters across iterations \citep{bottou2018optimization}. Since the number of subjects can vary considerably across providers, subjects from small providers are less likely to be selected than those from large providers under a simple random sampling scheme, possibly yielding insufficient updates on the effects of small providers and inflated variance of loss function gradients.

Responding to these challenges, this article introduces a fixed-effects approach augmented by neural networks for provider profiling. To the best of our knowledge, this approach represents a pioneering adaptation of deep learning technology to evaluate the performance of health care providers. The framework employs a generalized partially linear model (GPLM) incorporating a feedforward neural network (FNN) to capture nonlinear associations between risk factors and longitudinal outcomes, taking into account the variability in care quality across different providers. To ensure efficient implementation of the GPLM, we propose a novel stratified sampling-based stochastic optimization algorithm, which builds upon the widely-used AMSGrad algorithm in deep learning \citep{reddi2018amsgrad}. Given the substantial parameter space involved in training the deep learning profiling model, we incorporate two computational strategies: the curtailed training \citep{faraggi2001understanding} and ``dropout'' \citep{srivastava2014dropout}. As will be seen in due course, these strategies are designed to alleviate the issue of model overfitting. To identify providers exhibiting unusual performance, we offer a hypothesis testing procedure following the exact test-based approach \citep{wu2022improving}. This approach, distinct from methods relying on asymptotic approximations, offers methodological advantages for profiling small- to moderate-sized providers. Additionally, to facilitate the visualization and interpretation of profiling, we introduce exact test-based funnel plots \citep{spiegelhalter2005funnel, wu2023test} based on indirectly standardized quality metrics \citep{inskip1983methods}.

The remaining sections of this article are structured as follows: In \Cref{s:methods}, we present the GPLM that features neural networks, introduce the stratified sampling AMSGrad algorithm, outline the exact test for identifying providers with unusual performance, and elaborate on the corresponding funnel plots. \Cref{s:sim,s:app} illustrate the performance of our approach via simulation experiments and a data application involving Medicare ESRD beneficiaries undergoing kidney dialysis in the year 2020. \Cref{s:dis} offers a concluding discussion.

\section{Deep learning provider profiling}
\label{s:methods}

\subsection{Neural network model}
For $i = 1, \ldots{}, m$, let $n_i$ denote the number of subjects associated with provider $i$, where $m$ denotes the total number of providers. Let $n = \sum_{i=1}^m n_i$ be the total number of subjects. For $j = 1, \ldots{}, n_i$, let $Y_{ij}$ denote the outcome of subject $j$ with provider $i$, and let $\bZ_{ij}$ be a $p_0 \times 1$ vector of covariates for risk adjustment. We assume that the outcome $Y_{ij}$ satisfies the moment conditions $\Exp(Y_{ij}\mid\bZ_{ij}; \omega^*_{ij}) = \dot{h}(\omega^*_{ij})$ and $\Var(Y_{ij}\mid\bZ_{ij}; \omega^*_{ij}) = c(\phi)\ddot{h}(\omega^*_{ij})$ for known functions $c$ and $h$, where $\dot{h}(\omega^*_{ij})$ and $\ddot{h}(\omega^*_{ij})$ denote first- and second-order derivatives with respect to $\omega^*_{ij}$, respectively, and $\phi$ is a nuisance parameter. The specification of $c$ and $h$ typically depends on the type of outcome $Y_{ij}$. In this article, we focus on the commonly encountered continuous, binary, and Poisson outcomes, which correspond to the canonical identity, logit, and log link functions of $\dot{h}$, respectively. To associate the distribution of outcome $Y_{ij}$ with covariates $\bZ_{ij}$, we consider 
\begin{equation}
\label{eq:parlincom}
\omega^*_{ij} = \gamma_i + g^*(\bZ_{ij}),
\end{equation}
a partially linear predictor equal to the sum of a fixed provider effect $\gamma_i$ and an unknown real-valued function $g^*: \mathbb{R}^{p_0} \mapsto \mathbb{R}$ of risk factors $\bZ_{ij}$ that accounts for possible nonlinearity in covariate effects. Since it is theoretically known that a neural network with one hidden layer can approximate any continuous function if the number of nodes is sufficiently large \citep{fan2021selective}, the function $g^*$ will be represented by a neural network denoted as $g: \mathbb{R}^{p_0} \mapsto \mathbb{R}$, with output $g(\bZ_{ij})$ approximating the nonlinear component $g^*(\bZ_{ij})$ of $\omega^*_{ij}$.

\begin{figure}[hbtp]
\centering
\includegraphics[width=0.6\textwidth]{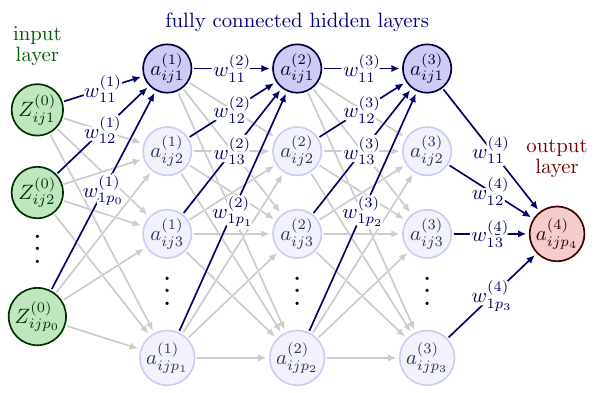}
\caption{A feedforward neural network with three fully connected hidden layers for subject $j$ of provider $i$ with input vector $\ba^{(0)}_{ij} = \bZ_{ij} = [Z_{ij1}, \ldots{}, Z_{ijp_0}]^\top$. Other layers are defined recursively by the formula $\bm{a}_i^{(l+1)} = g_{l+1}(\mathbf{w}^{(l+1)}\bm{a}_i^{(l)} + \mathbf{b}^{(l+1)})$, $l = 0, \ldots{}, 3$, where $\mathbf{w}^{(l+1)}$ is a weight matrix and $\mathbf{b}^{(l+1)}$ is a bias vector. Information transmission to the first node in hidden and output layers is highlighted with weights.}
\label{fig:FNN}
\end{figure}

We consider an FNN $g$ with $L$ hidden layers, where $L \in \{0\} \cup \mathbb{N}$. For subject $j$ from provider $i$, let $l \in \{0, \ldots{}, L + 1\}$ index the $l$th layer (input layer if $l = 0$ and output layer if $l = L + 1$), and let $\ba_{ij}^{(0)} = \bZ_{ij}$ be the input vector. Subsequent layers can be determined recursively via the following alternating affine and nonlinear transformations:
\begin{equation}
\label{eq:FNNrecursion}
\ba_{ij}^{(l+1)} = g_{l+1}(\bw^{(l+1)} \ba_{ij}^{(l)} + \bb^{(l+1)}), \quad l = 0, \ldots{}, L,
\end{equation}
where $\ba_{ij}^{(l+1)}$ is a $p_{l+1}$-dimensional output vector from a prespecified activation function $g_{l+1}$ applied element-wise to $\bw^{(l+1)} \ba_{ij}^{(l)}$, $\bw^{(l+1)}$ is an $p_{l+1} \times p_l$ weight matrix whose $k$th row is denoted as $\bw^{(l+1)}_k$, and $\bb^{(l+1)}$ is an $p_{l+1}$-dimensional bias vector whose $k$th element is denoted as $b^{(l+1)}_k$. Subjects from all providers share the same set of unknown parameters $\bw^{(l+1)}$ and $\bb^{(l+1)}$. The output $a_{ij}^{(L+1)} = g(\bZ_{ij}; \bw, \bb)$ of this FNN being a scalar indicates that $p_{L+1} = 1$. Let $\btheta = [\bgamma^\top, \bw^\top, \bb^\top]^\top$ be a vector of all parameters in the FNN, where $\bgamma = [\gamma_1, \ldots{}, \gamma_m]^\top$, $\bw$ is a vectorization of $\bw^{(1)}$, $\ldots$, $\bw^{(L+1)}$, and $\bb$ is a vectorization of $\bb^{(1)}$, $\ldots$, $\bb^{(L+1)}$. The motivation for constructing the FNN is to accurately quantify effects $\bgamma$ and profile providers while reducing potential biases induced by complex confounding of risk factors. As an illustration, \Cref{fig:FNN} describes an FNN with three fully connected hidden layers.

\subsection{Stochastic optimization with stratified sampling}
\label{s:stoopt}
Training neural network models can pose significant challenges in practice. When a large volume of data are involved, it is often computationally overwhelming to use either the second-order conditions of a loss function or gradient information from all observations across all iterations of an algorithm. Therefore, gradient-based stochastic optimization methods are preferred rather than conventional Newton-type or gradient descent methods. Among the myriad of deep learning methods \citep{ruder2016overview}, in this article, we consider AMSGrad \citep{reddi2018amsgrad}, a state-of-the-art approach that overcomes the convergence pitfalls of the adaptive moment estimation \citep[Adam,][]{kingma2015adam}. To reduce the variance of parameter gradients in the presence of provider-level clustering, we additionally incorporate a stratified sampling mechanism into the AMSGrad. Rather than sampling observations uniformly in the training set, we draw a fixed proportion of observations associated with each provider. The resulting \Cref{alg:SSAMSGrad} is thus termed stratified sampling AMSGrad (SSAMSGrad), which was inspired by but different from studies leveraging SGD with stratified sampling \citep{liu2022parameter}. 

Given observations $\{(Y_{ij}, \bZ_{ij}): (i, j) \in \cD\}$, where $\cD = \{(i, j): i = 1, \ldots{}, m,\,\, j = 1, \ldots{}, n_i\}$, we define the loss function $-\ell(\btheta)$ to be the following:
\begin{equation}
\label{eq:llkd}
-\ell(\btheta) = -
\sum_{i=1}^{m} \sum_{j=1}^{n_i} \ell_{ij}(\btheta)
= -\sum_{i=1}^{m} \sum_{j=1}^{n_i} [Y_{ij}\omega_{ij} - h(\omega_{ij})],
\end{equation}
where $\omega_{ij} = \gamma_i + g(\bZ_{ij}; \bw, \bb)$. This loss function is derived from the fact that
\[
-\frac{\partial\ell(\btheta)}{\partial \omega_{ij}} \cdot \frac{1}{c(\phi)} = \frac{\partial\Exp(Y_{ij} \mid \bZ_{ij}; \omega_{ij})}{\partial \omega_{ij} } \cdot \frac{Y_{ij} - \Exp(Y_{ij} \mid \bZ_{ij}; \omega_{ij})}{\Var(Y_{ij} \mid \bZ_{ij}; \omega_{ij})}
\]
is the neural network representation of the generalized estimating function. Next, $\cD$ is randomly split into a training set $\cT$ and a validation set $\cV$ via stratified sampling so that $|\cT|$, the number of observations in $\cT$, is equal to $\lfloor \delta |\cD| \rfloor$ with a prespecified proportion $\delta \in (0.5, 1)$. The weights $\bw$ are initialized following the Glorot uniform initialization \citep{glorot2010understanding}, while biases $\bb$ and provider effects $\bgamma$ are initialized at $\mathbf{0}$. At iteration $s \in \mathbb{N}$, the stochastic gradient is $\bg_{(s)} = - |\cT_{(s)}|^{-1} \sum_{(i, j) \in \cT_{(s)}} \dot{\ell}_{ij}(\btheta_{(s-1)})$, where $\dot{\ell}_{ij}(\btheta_{(s-1)})$ is calculated by the chain rule (details in Appendix A of the Supplementary Material), and $|\cT_{(s)}| = \lfloor \xi |\cT| \rfloor$ with $\xi \in (0, 1)$ being the sampling proportion. Unlike the conventional AMSGrad, here $\cT_{(s)}$ is formed by stratified sampling observations across providers using $\xi$. Further, the step size $\br_{(s)} / (\sqrt{\hat \bv_{(s)}} + \epsilon)$ is determined as the moving average of the updated stochastic gradient $\bg_{(s)}$ and the past unnormalized step size $\br_{(s-1)}$, then normalized by $\sqrt{\hat \bv_{(s)}} + \epsilon$, with $\epsilon$ being a very small positive number (e.g., $10^{-8}$) that ensures a nonzero denominator. To obtain a non-decreasing sequence of normalizers, $\hat \bv_{(s)}$ is defined as the maximum of $\bv_{(s)}$ and $\hat \bv_{(s-1)}$, where $\bv_{(s)}$ is the moving average of the element-wise square of $\bg_{(s)}$ and the past copy $\bv_{(s-1)}$. All parameters $\btheta_{(s-1)}$ are updated by subtracting $\eta_{(s)}\br_{(s)} / (\sqrt{\hat \bv_{(s)}} + \epsilon)$, where $\eta_{(s)} > 0$ is a learning rate that is allowed to decay across iterations.

\begin{algorithm}[htbp]
\SetAlCapFnt{\small}
\SetAlgoLined
\SetKwRepeat{Do}{do}{while}
\SetKw{KwFrom}{from}
initialize $s = 0$, $\psi \gg 10^8$, $\bgamma_{(0)} = \mathbf{0}$, $\bb_{(0)} = \mathbf{0}$, and $\br_{(0)} = \hat \bv_{(0)} = \bv_{(0)} = \mathbf{0}$\;
\For{$l$ \KwFrom $0$ \KwTo $L+1$}{
$\bw^{(l+1)}_{(0)} \sim \mathrm{Uniform}(-\sqrt{6}/\sqrt{p_l + p_{l+1}}, \sqrt{6}/\sqrt{p_l + p_{l+1}})$\;
}
set $\delta \in (0.5, 1)$, $\eta = 10^{-3}$, $\xi \in (0, 1)$, $\beta_1 = 0.9$, $\beta_2 = 0.999$, $\epsilon = 10^{-8}$, and $u \in \mathbb{N}$\;
split $\cD$ into a training set $\cT$ and a validation set $\cV$ via stratified sampling with $|\cT| = \lfloor \delta |\cD| \rfloor$\;
\Do{$-|\cV|^{-1} \sum_{(i, j) \in \cV} \ell_{ij}(\btheta_{(s)}) > \psi$ {\rm across at most $u-1$ consecutive iterations}}{
$s \leftarrow s + 1$\;
$\bg_{(s)} = - |\cT_{(s)}|^{-1} \sum_{(i, j) \in \cT_{(s)}} \dot{\ell}_{ij}(\btheta_{(s-1)})$, where $\cT_{(s)}$ is a stratified random sample of $\cT$ with $|\cT_{(s)}| = \lfloor \xi |\cT| \rfloor$\;
$\br_{(s)} = \beta_1 \br_{(s-1)} + (1-\beta_1)\bg_{(s)}$\;
$\bv_{(s)} = \beta_2 \bv_{(s-1)} + (1-\beta_2) \bg_{(s)} \odot \bg_{(s)}$, where $\odot$ denotes the element-wise product\;
$\hat \bv_{(s)} = \max\{\bv_{(s)}, \hat \bv_{(s-1)}\}$\;
$\eta_{(s)} = \eta / \sqrt{s}$\;
$\btheta_{(s)} = \btheta_{(s-1)} - \eta_{(s)} \br_{(s)} / (\sqrt{\hat \bv_{(s)}} + \epsilon)$, where square root and division are element-wise\;
$\psi \leftarrow \min\{\psi, -|\cV|^{-1} \sum_{(i, j) \in \cV} \ell_{ij}(\btheta_{(s)})\}$\;
}
\caption{SSAMSGrad}
\label{alg:SSAMSGrad}
\end{algorithm}

As remedies for model overfitting, we consider two standard strategies, curtailed training (early stopping) \citep{faraggi2001understanding} and ``dropout'' \citep{srivastava2014dropout}. To incorporate early stopping in the SSAMSGrad, we track the sequence of loss function values on the validation set $\cV$; the algorithm is terminated when the validation loss is higher than the running minimum $\psi$ across $u$ or more consecutive iterations, where $u \in \mathbb{N}$ is a prespecified natural number (e.g., $5$). Dropout is a technique that randomly drops out nodes in a neural network (illustrated in Figure 1 of the Supplementary Material). To compute $\dot{g}(\bZ_{ij}; \bw_{(s-1)}, \bb_{(s-1)})$ for $(i, j) \in \cT_{(s)}$ at iteration $s$, each node in input and hidden layers is subject to temporary removal from the network with a retention probability $\upsilon$ (usually closer to 1 than to 0.5) independent of other nodes. However, when computing $\ell_{ij}(\btheta_{(s)})$ for $(i, j) \in \cV$, all nodes are kept in the network without dropout, but the outgoing weights of a dropout node in the training stage will be multiplied by $\upsilon$.

\subsection{Identifying outlying providers}
\label{subs:exacttest}
As noted earlier, the overarching goal of provider profiling is to identify providers having subpar performance with respect to a predefined standard or benchmark. A principled approach to identifying these providers is to derive a provider-specific hypothesis testing procedure. Here, the null hypothesis can be written as $H_{0,i}: \gamma_i = f(\bgamma)$, where $f: \mathbb{R}^m \rightarrow \mathbb{R}$ is a real-valued deterministic function dictated by an entity accountable for health care regulation and oversight. Since $\bgamma$ is unobserved in practice, $\bgamma$ is often replaced by its estimate $\hat\bgamma$. A popular candidate of $f$ in the profiling literature is the median \citep{he2013evaluating, estes2018time, estes2020profiling, wu2022analysis, wu2022improving, wu2023test}, a more robust measure compared to the mean. In this case, the hypothetical provider with the median provider effect is called the population norm. Since $f(\hat\bgamma)$ is a very accurate estimate of $f(\bgamma)$ in most profiling applications with large-scale data, we hereafter do not distinguish between $f(\bgamma)$ and $f(\hat\bgamma)$.

A recent profiling study under the framework of generalized linear models (GLMs) suggests that distribution-based exact tests tend to have controlled type I error and improved statistical power compared with score and Wald tests, especially when numerous providers have a small number of subjects or limited variation in the outcome \citep{wu2022improving}. Here we extend the exact-test-based profiling approach to GPLMs with FNNs. Since constructing the exact test requires positing a model for the conditional distribution of the outcome $Y_{ij}$ as opposed to only specifying the expectation and variance, now we make a simplifying assumption that $Y_{ij}$ follows a distribution in the exponential family given $\bZ_{ij}$, $\omega_{ij}$, and $\phi$, i.e.,
\begin{equation}
\label{eq:expfam}
\pi(Y_{ij}\mid\bZ_{ij}; \omega_{ij}, \phi)
\propto \exp\left\{\frac{Y_{ij}\omega_{ij} - h(\omega_{ij})}{c(\phi)}\right\},
\end{equation}
where $\omega_{ij} = \gamma_i + g(\bZ_{ij}; \bw, \bb)$. Observe that the outcomes $\{Y_{ij}: j =1, \ldots{}, n_i\}$ from provider $i$ are independent given risk factors $\bZ_i = [\bZ^\top_{i1}, \ldots{}, \bZ^\top_{in_i}]^\top$ and the provider effect $\gamma_i$. Therefore, we can derive the exact test under the null hypothesis $H_{0,i}$ leveraging the conditional distribution of $O_i \coloneqq \sum_{j=1}^{n_i} Y_{ij}$ given $\bZ_i$. Since training the FNN involves a large number of subjects according to \eqref{eq:llkd}, we make another assumption that $g(\bZ_{ij}; \bw, \bb)$ can be well approximated by $g(\bZ_{ij}; \hat\bw, \hat\bb)$, where $\hat\bw$ and $\hat\bb$ are estimates of weights $\bw$ and biases $\bb$, respectively. Similar treatments have been considered in previous studies on profiling methods \citep{he2013evaluating, estes2018time, estes2020profiling, xia2022accounting, wu2022improving, wu2023test}. In what follows, we derive the cumulative distribution function (CDF) of $O_i$ given $\bZ_i$ for three common outcome types for $Y_{ij}$.

If $Y_{ij}$ is Gaussian distributed with nuisance variance $\sigma^2$, then $O_i \mid \bZ_i \sim \cN(\sum_{j=1}^{n_i} \dot{h}\{\gamma_i + g(\bZ_{ij})\}, n_i\sigma^2)$, where $\sigma^2$ can be substituted with its unbiased estimator $\hat\sigma^2 = (n - m - p_0)^{-1}\sum_{i=1}^m \sum_{j=1}^{n_i} \{Y_{ij} - \hat\gamma_i - g(\bZ_{ij}; \hat\bw, \hat\bb)\}^2$, an unbiased estimator of $\sigma^2$. For $o \in \mathbb{R}$, the CDF of $O_i$ conditional on $\bZ_i$ is given by
\[
F_i(o \mid \bZ_i; \gamma_i, \bw, \bb)
= \frac{1}{\sqrt{2\pi n_i\hat\sigma^2}} \int^o_{-\infty}  \exp\left\{-\frac{1}{2n_i\hat\sigma^2}\left[x - \sum_{j=1}^{n_i} \dot{h}\{\gamma_i + g(\bZ_{ij}; \bw, \bb)\}\right]^2\right\}\,\rd x.
\]

If $Y_{ij}$ follows a Bernoulli distribution, we have $Y_{ij} \mid \bZ_i \sim \mathrm{Bernoulli}(\dot{h}\{\gamma_i + g(\bZ_{ij}; \bw, \bb)\})$. It follows that $O_i \mid \bZ_i$ has a Poisson-binomial distribution \citep{chen1997statistical, johnson2005univariate}. Let $S_i \coloneqq \{1, \ldots{}, n_i\}$, $\mathcal{A}_{il} \coloneqq \{A_i \subset S_i : |A_i| = l\}$, and $A_i^c \coloneqq S_i \setminus A_i$. For $o \in \{0\} \cup S_i$, the CDF of $O_i$ given $\bZ_i$ is
\begin{equation}
\label{eq:pbcdf}
F_i(o \mid \bZ_i; \gamma_i, \bw, \bb)
= \sum_{l=0}^o \sum_{A_i \in \mathcal{A}_{il}} 
\left\{\prod_{a \in A_i} \dot{h}\{\gamma_i + g(\bZ_{ia}; \bw, \bb)\} \prod_{q \in A_i^c} [1 - \dot{h}\{\gamma_i + g(\bZ_{iq}; \bw, \bb)\}]\right\},
\end{equation}
where we follow the convention that an empty product equals one.

If $Y_{ij}$ follows a Poisson distribution, i.e., $Y_{ij} \mid \bZ_i \sim \mathrm{Poisson}(\dot{h}\{\gamma_i + g(\bZ_{ij}; \bw, \bb)\})$, then $O_i \mid \bZ_i \sim \mathrm{Poisson}(\sum_{j=1}^{n_i} \dot{h}\{\gamma_i + g(\bZ_{ij}; \bw, \bb)\})$. For $o \in \{0\} \cup \mathbb{N}$, the CDF of $O_i$ conditional on $\bZ_i$ is
\[
F_i(o \mid \bZ_i; \gamma_i, \bw, \bb)
= \frac{1}{o!}\left[\sum_{j=1}^{n_i} \dot{h}\{\gamma_i + g(\bZ_{ij}; \bw, \bb)\}\right]^{o}\exp\left\{-\sum_{j=1}^{n_i} \dot{h}\{\gamma_i + g(\bZ_{ij}; \bw, \bb)\}\right\}.
\]

With the CDFs, the mid $p$-value $P_i$ for a two-sided exact test against the null hypothesis $H_{0,i}$ is
\begin{equation}
\label{eq:exactpvalue}
P_i = 2 \cdot \min\{G_i(O_i \mid \bZ_i; f(\hat\bgamma), \hat\bw, \hat\bb), 1 - G_i(O_i \mid \bZ_i; f(\hat\bgamma), \hat\bw, \hat\bb)\},
\end{equation}
where $G_i(o \mid \bZ_i; f(\bgamma), \bw, \bb) = F_i(o \mid \bZ_i; f(\bgamma), \bw, \bb) - 0.5\Pr(O_i = o \mid \bZ_i; f(\bgamma), \bw, \bb)$ is termed the sub-CDF of $F_i(o \mid \bZ_i; f(\bgamma), \bw, \bb)$. For any $\alpha \in (0,1)$, the lower limit $\ubar{\gamma}_i$ and upper limit $\overline{\gamma}_i$ of a $100(1-\alpha)\%$ confidence interval of a provider effect $\gamma_i$ are determined by equations $G_i(O_i \mid \bZ_i; \ubar{\gamma}_i, \hat\bw, \hat\bb) = 1 - \alpha_1$ and $G_i(O_i \mid \bZ_i; \overline{\gamma}_i, \hat\bw, \hat\bb) = \alpha_2$, respectively, where $\alpha_1, \alpha_2 \in [0,1)$ with $\alpha_1 + \alpha_2 = \alpha$.

\subsection{Visualizing provider profiling}
Funnel plots, originally designed as a graphical tool for meta-analysis, have gained popularity for institutional comparison. This is primarily attributed to their interpretability in effectively identifying providers with outstanding performance based on patient-centered outcomes \citep{spiegelhalter2012statistical, tang2020constructing}. Similar to \citet{wu2023test}, we have developed a customized funnel plot that is specifically designed to facilitate the visualization of provider profiling using the proposed exact test.

A funnel plot generally consists of four components: a standardized measure of interest, a target $\tau$ of the measure, the precision of the measure, and control limits specific to a $p$-value $\alpha \in (0, 1)$. Although not without controversy \citep{george2017mortality}, indirect standardization is a widely utilized approach in epidemiology and provider profiling that compares the observed number of events in a specific group with the expected number of events in a reference population \citep{he2013evaluating, estes2018time, wu2022analysis}. By quantifying the deviation from the expected outcome level, indirect standardization enables the identification of whether the observed number of outcomes within a specific provider is more or fewer than expected. Moreover, this approach, which takes into account the expected number of events based on a reference population, offers numerically stable standardized metrics, particularly when evaluating relatively small providers \citep{inskip1983methods}. We defer the discussion of the pros and cons of indirect standardization to \Cref{s:dis}.

Under the current FNN framework, an indirectly standardized ratio for provider $i$ can be defined as
\begin{equation}
\label{eq:measure}
T_i \coloneqq \frac{O_i}{E_i} = \frac{\sum_{j=1}^{n_i} Y_{ij}}{\sum_{j=1}^{n_i} \dot{h}\{f(\hat\bgamma) + g(\bZ_{ij}; \hat\bw, \hat\bb)\}},    
\end{equation}
where $O_i$ denotes the sum of observed outcomes for provider $i$, and $E_i$ denotes the sum of expected outcomes with the provider effect set equal to $f(\hat\bgamma)$. When the ratio $T_i$ is less than one, it indicates that the observed outcomes within a specific group are lower than expected, based on a predetermined reference population. Conversely, if the ratio is greater than one, it signifies that the observed outcomes are higher than expected. The value of one is often chosen as the target for an indirectly standardized ratio due to its intuitive interpretation. However, in certain applications, a different value $\tau$ may also be of interest. As noted in \citet{spiegelhalter2005funnel}, a general target $\tau$ implies that for an in-control (non-outlying) provider $i$, the CDF of $O_i$ given $\bZ_i$ is $F_i(o \mid \bZ_i; f(\hat\bgamma), \hat\bw, \hat\bb, \tau)$, a modification of $F_i(o \mid \bZ_i; f(\hat\bgamma), \hat\bw, \hat\bb)$ whose $\dot{h}$'s are multiplied by $\tau$. In this case, the precision $\rho_i(\tau)$ of $T_i$ is simply $E_i^2 / V_i(\tau)$, where
\begin{subnumcases}
{\label{generalcase} V_i(\tau) =}
n_i \hat\sigma^2 & if $Y_{ij}$ is Gaussian distributed, \nonumber \\ 
\tau\sum_{j=1}^{n_i} \dot{h}\{f(\hat\bgamma) + g(\bZ_{ij}; \hat\bw, \hat\bb)\} [1 - \tau\dot{h}\{f(\hat\bgamma) + g(\bZ_{ij}; \hat\bw, \hat\bb)\}] & if $Y_{ij}$ is Bernoulli distributed, \nonumber \\
\tau\sum_{j=1}^{n_i} \dot{h}\{f(\hat\bgamma) + g(\bZ_{ij}; \hat\bw, \hat\bb)\} & if $Y_{ij}$ is Poisson distributed, \nonumber
\end{subnumcases}
is the variance of $O_i$ given $\bZ_i$ with $F_i(o \mid \bZ_i; f(\hat\bgamma), \hat\bw, \hat\bb, \tau)$ as the CDF. As the definition suggests, the precision $\rho_i(\tau)$ can be interpreted as the inverse of the squared coefficient of variation for the distribution of $O_i$ given $\bZ_i$ and then divided by $\tau^2$ (i.e., $\{V_i(\tau)/(\tau E_i)^2\}^{-1}/\tau^2$).

Since the outcome $Y_{ij}$ can be discrete, we adopt an interpolation approach to establish the control limits of $T_i$ given $p$-value $\alpha$ and target $\tau$ \citep{spiegelhalter2005funnel}. Let $\tilde{O}_i(\alpha, \tau) = \inf\{o:G_i(O_i \mid \bZ_i; f(\hat\bgamma), \hat\bw, \hat\bb, \tau) \geq \alpha\}$, where $G_i(o \mid \bZ_i; f(\hat\bgamma), \hat\bw, \hat\bb, \tau)$ is the sub-CDF of $F_i(o \mid \bZ_i; f(\hat\bgamma), \hat\bw, \hat\bb, \tau)$ as defined in \Cref{subs:exacttest}. Let the interpolation weight $\lambda_i(\alpha, \tau)$ be
\[
\lambda_i(\alpha, \tau) = \inf\{\lambda \in [0,1]: \lambda G_i(\tilde{O}^-_i(\alpha, \tau); f(\hat\bgamma), \hat\bw, \hat\bb, \tau) + (1-\lambda)G_i(\tilde{O}_i(\alpha, \tau); f(\hat\bgamma), \hat\bw, \hat\bb, \tau) = \alpha\},
\]
where $G_i(o^{-}; f(\hat\bgamma), \hat\bw, \hat\bb, \tau) = \lim_{o^* \rightarrow o^{-}} G_i(o^*; f(\hat\bgamma), \hat\bw, \hat\bb, \tau)$. Let $O_i(\alpha, \tau) = \tilde{O}_i(\alpha, \tau) - \lambda_i(\alpha, \tau)$. Then the interpolated control limits of $T_i$ for $p$-value $\alpha$ and target $\tau$ are
\begin{equation}
\label{eq:ctrllimits}
[T_i(\alpha_1, \tau), T_i(1-\alpha_2, \tau)] =
\left[\frac{O_i(\alpha_1, \tau)}{E_i}, \frac{O_i(1-\alpha_2, \tau)}{E_i}\right],
\end{equation}
where $\alpha_1, \alpha_2 \in [0,1)$ with $\alpha_1 + \alpha_2 = \alpha$.

\section{Simulation experiments}
\label{s:sim}
We perform simulation analyses to evaluate the proposed profiling methods augmented by neural networks. Since the outcomes for Gaussian and Poisson distributions have been well-studied in the literature, we will primarily focus on Bernoulli outcomes throughout the remainder of this article.

\subsection{Comparing GPLM and GLM}

In the first experiment, our aim is to compare the neural-network-based GPLM with GLM in terms of predictive power. To this end, we consider the following data-generating mechanism:
\begin{itemize}
\item The number of providers $m$ is set to 100, 300, or 500;
\item Provider-specific subject counts $\{n_i: i = 1, \ldots, m\}$ are drawn from a Poisson distribution with the mean $\nu$ equal to 50, 100, or 200; to preclude very small providers, a subject count is truncated to be at least 20;
\item Provider effects $\bgamma$ are sampled from a Gaussian distribution $\mathcal{N}(\mu, \sigma^2)$ with $\mu = \log(4/11)$ and $\sigma = 0.4$, and are fixed throughout all simulated data sets;
\item Following \citet{kalbfleisch2013monitoring}, subject-specific covariates $\bZ_{ij}$ are generated according to
\begin{equation}
\label{eq:samplecovar}
\bZ_{ij} \sim \mathcal{N}\left((\rho/\sigma)(\gamma_i - \mu)\be,
\mathbf{\Omega} - \rho^2 \mathbf{J}
\right),\, j = 1, \ldots{}, n_i,
\end{equation}
where $\rho = 0$ and 0.5, respectively, $\mathbf{\Omega}$ is a $3 \times 3$ matrix with diagonal ones and off-diagonal $\rho$'s, $\be$ is a vector of 3 ones, and $\mathbf{J}$ is a $3 \times 3$ matrix of ones; consequently, $\mathrm{Corr}(\bZ_{ij}, \gamma_i) = \rho\be$ and $\bZ_{ij} \sim \mathcal{N}(\mathbf{0}, \mathbf{\Omega})$;
\item A function of linear associations is specified as
\begin{equation}
\label{eq:linear}
g_1^*(\bZ_{ij}) = Z_{ij1} + 0.5Z_{ij2} - Z_{ij3};
\end{equation}
\item A function of linear and nonlinear associations is specified as 
\begin{equation}
\label{eq:nonlinear}
g_2^*(\bZ_{ij}) = Z_{ij1} + 0.5Z_{ij2} - Z_{ij3} + 0.2 Z_{ij1}Z_{ij2} + 0.8Z^2_{ij2} + 0.4 \cos(Z_{ij1}) \sin(Z_{ij3});
\end{equation}
\item The outcome $Y_{ij}$ is sampled from a Bernoulli distribution with the mean equal to $\expit\{\gamma_i + g_k^*(\bZ_{ij})\}$, where $\expit$ denotes the logistic function and $k = 1, 2$; and
\item In each scenario, 500 simulated data sets are generated.
\end{itemize}
We fit fixed-effect GPLM and GLM \citep{wu2022scalable} to each simulated data set to compare the predictive ability of the two models, where the GPLM is implemented as SSAMSGrad in \Cref{alg:SSAMSGrad}. All FNNs include an input layer of 3 nodes, two hidden layers of 32 and 16 nodes, and an output layer of 1 node. The corresponding activation functions are the rectified linear unit (ReLU), ReLU, and identity, respectively. Given the two sets of predicted probabilities, we calculate the accuracy (the sum of true positives and true negatives divided by the number of subjects), sensitivity (also known as recall, the proportion of true positives among all actual positives), specificity (the proportion of true negatives among all actual negatives), precision (the proportion of true positives among all predicted positives), F1 (the harmonic mean of sensitivity and precision), and the area under the receiver operating characteristic curve (AUC). A higher value of a metric indicates better performance.

\Cref{tab:linearvsnonlinear} presents the mean and standard deviation of all metrics for the GPLM and GLM with varied $m$, $\rho$, and $\nu$ for linear \eqref{eq:linear} and nonlinear model \eqref{eq:nonlinear}. Holding other things constant, an increase in $m$ or $\nu$ leads to lower specificity but a higher value in the remaining five metrics; an increase in $\rho$ leads to a lower value in all performance metrics. When the true model is linear (Panel A), GPLM and GLM have similar performance metrics for $\nu = 50$; when $\nu = 100$ and $200$, GLM slightly outperforms GPLM in all criteria except sensitivity for $m = 300$ and $500$. When the true model is nonlinear (Panel B), the GPLM consistently outperforms the GLM across all performance metrics for varied $m$, $\rho$, and $\nu$. As expected, simulation experiments in \Cref{tab:linearvsnonlinear} demonstrate that the GPLM with an FNN  excels in characterizing complex associations between the outcome and covariates.

\begin{sidewaystable}[htbp]
\caption{Performance of the generalized partially linear model (GPLM) and generalized linear model (GLM). The GPLM is implemented as the SSAMSGrad. In each setting, the mean and standard deviation (in parentheses) of each metric are derived from 500 simulated data sets.\label{tab:linearvsnonlinear}}
\centering
\resizebox{\columnwidth}{!}{%
\begin{tabular}{cccccccccccccc}
\toprule
\toprule
\multicolumn{14}{c}{Panel A: $g^*_1(\bZ_{ij})$ as in \eqref{eq:linear}} \\
\midrule
\multirow{3}{*}{$m$} & \multirow{3}{*}{metric} & \multicolumn{6}{c}{GPLM} & \multicolumn{6}{c}{GLM} \\
\cmidrule(r){3-8} \cmidrule(l){9-14}
& & \multicolumn{3}{c}{$\rho=0$} & \multicolumn{3}{c}{$\rho=0.5$} & \multicolumn{3}{c}{$\rho=0$} & \multicolumn{3}{c}{$\rho=0.5$} \\
\cmidrule(r){3-5} \cmidrule(lr){6-8} \cmidrule(lr){9-11} \cmidrule(l){12-14}
& & $\nu=50$ & $\nu=100$ & $\nu=200$ & $\nu=50$ & $\nu=100$ & $\nu=200$ & $\nu=50$ & $\nu=100$ & $\nu=200$ & $\nu=50$ & $\nu=100$ & $\nu=200$ \\
\midrule
\multirow{6}{*}{100} 
 & accuracy & 0.762 (0.014) & 0.765 (0.010) & 0.766 (0.007) & 0.746 (0.015) & 0.747 (0.011) & 0.748 (0.008) & 0.761 (0.014) & 0.767 (0.009) & 0.768 (0.007) & 0.745 (0.015) & 0.748 (0.011) & 0.750 (0.008) \\
 & sensitivity & 0.870 (0.018) & 0.876 (0.014) & 0.879 (0.010) & 0.881 (0.020) & 0.885 (0.015) & 0.889 (0.012) & 0.871 (0.015) & 0.877 (0.011) & 0.879 (0.008) & 0.880 (0.017) & 0.887 (0.013) & 0.890 (0.011)\\
 & specificity & 0.543 (0.035) & 0.540 (0.028) & 0.538 (0.021) & 0.452 (0.043) & 0.447 (0.034) & 0.443 (0.027) & 0.537 (0.030) & 0.542 (0.024) & 0.541 (0.017) & 0.448 (0.037) & 0.447 (0.030) & 0.445 (0.026)\\
 & precision & 0.794 (0.016) & 0.795 (0.012) & 0.794 (0.008) & 0.778 (0.017) & 0.777 (0.013) & 0.776 (0.009) & 0.792 (0.015) & 0.796 (0.011) & 0.796 (0.008) & 0.777 (0.016) & 0.777 (0.013) & 0.777 (0.009)\\
 & F1 & 0.830 (0.012) & 0.833 (0.008) & 0.834 (0.006) & 0.826 (0.013) & 0.827 (0.010) & 0.829 (0.008) & 0.830 (0.011) & 0.834 (0.008) & 0.835 (0.006) & 0.825 (0.013) & 0.828 (0.010) & 0.829 (0.008)\\
 & AUC & 0.811 (0.014) & 0.815 (0.010) & 0.817 (0.007) & 0.775 (0.017) & 0.777 (0.012) & 0.779 (0.009) & 0.809 (0.014) & 0.816 (0.010) & 0.819 (0.007) & 0.771 (0.017) & 0.778 (0.012) & 0.781 (0.009)\\
\midrule
\multirow{6}{*}{300} 
 & accuracy & 0.764 (0.008) & 0.765 (0.006) & 0.767 (0.004) & 0.747 (0.009) & 0.748 (0.007) & 0.748 (0.005) & 0.762 (0.008) & 0.766 (0.005) & 0.768 (0.004) & 0.745 (0.009) & 0.748 (0.007) & 0.750 (0.005) \\
 & sensitivity & 0.876 (0.010) & 0.879 (0.007) & 0.882 (0.005) & 0.886 (0.011) & 0.889 (0.009) & 0.892 (0.007) & 0.871 (0.009) & 0.876 (0.006) & 0.880 (0.005) & 0.881 (0.010) & 0.887 (0.008) & 0.890 (0.007) \\
 & specificity & 0.535 (0.022) & 0.534 (0.016) & 0.534 (0.011) & 0.446 (0.025) & 0.441 (0.020) & 0.438 (0.016) & 0.540 (0.019) & 0.541 (0.013) & 0.540 (0.010) & 0.449 (0.022) & 0.447 (0.017) & 0.445 (0.016) \\
 & precision & 0.793 (0.009) & 0.793 (0.007) & 0.793 (0.005) & 0.776 (0.010) & 0.775 (0.007) & 0.775 (0.006) & 0.794 (0.009) & 0.795 (0.006) & 0.795 (0.004) & 0.776 (0.010) & 0.777 (0.007) & 0.777 (0.005) \\
 & F1 & 0.832 (0.007) & 0.834 (0.005) & 0.835 (0.004) & 0.827 (0.007) & 0.828 (0.006) & 0.829 (0.005) & 0.831 (0.007) & 0.834 (0.005) & 0.835 (0.004) & 0.825 (0.008) & 0.828 (0.006) & 0.829 (0.005) \\
 & AUC & 0.813 (0.009) & 0.815 (0.006) & 0.817 (0.004) & 0.776 (0.009) & 0.778 (0.007) & 0.779 (0.005) & 0.810 (0.009) & 0.815 (0.006) & 0.818 (0.004) & 0.772 (0.010) & 0.778 (0.007) & 0.781 (0.005) \\
\midrule
\multirow{6}{*}{500}
 & accuracy & 0.764 (0.006) & 0.765 (0.004) & 0.767 (0.003) & 0.747 (0.006) & 0.748 (0.005) & 0.749 (0.004) & 0.762 (0.006) & 0.765 (0.004) & 0.768 (0.003) & 0.744 (0.006) & 0.748 (0.005) & 0.750 (0.004) \\
 & sensitivity & 0.878 (0.008) & 0.881 (0.006) & 0.882 (0.004) & 0.888 (0.009) & 0.891 (0.007) & 0.893 (0.006) & 0.871 (0.007) & 0.877 (0.005) & 0.880 (0.004) & 0.880 (0.008) & 0.887 (0.006) & 0.890 (0.005) \\
 & specificity & 0.532 (0.018) & 0.531 (0.012) & 0.533 (0.009) & 0.439 (0.020) & 0.437 (0.015) & 0.437 (0.012) & 0.539 (0.014) & 0.540 (0.010) & 0.541 (0.008) & 0.448 (0.017) & 0.446 (0.013) & 0.446 (0.011) \\
 & precision & 0.792 (0.008) & 0.792 (0.005) & 0.793 (0.004) & 0.775 (0.007) & 0.774 (0.006) & 0.775 (0.004) & 0.793 (0.007) & 0.794 (0.005) & 0.796 (0.003) & 0.776 (0.007) & 0.776 (0.005) & 0.777 (0.004) \\
 & F1 & 0.833 (0.005) & 0.834 (0.004) & 0.835 (0.003) & 0.828 (0.005) & 0.828 (0.005) & 0.829 (0.004) & 0.831 (0.005) & 0.834 (0.004) & 0.836 (0.003) & 0.825 (0.005) & 0.828 (0.004) & 0.830 (0.004) \\
 & AUC & 0.813 (0.006) & 0.815 (0.005) & 0.818 (0.003) & 0.776 (0.007) & 0.778 (0.005) & 0.780 (0.004) & 0.810 (0.006) & 0.815 (0.004) & 0.819 (0.003) & 0.771 (0.007) & 0.778 (0.005) & 0.781 (0.004) \\
\toprule
\multicolumn{14}{c}{Panel B: $g^*_2(\bZ_{ij})$ as in \eqref{eq:nonlinear}} \\
\midrule
\multirow{3}{*}{$m$} & \multirow{3}{*}{metric} & \multicolumn{6}{c}{GPLM} & \multicolumn{6}{c}{GLM} \\
\cmidrule(r){3-8} \cmidrule(l){9-14}
& & \multicolumn{3}{c}{$\rho=0$} & \multicolumn{3}{c}{$\rho=0.5$} & \multicolumn{3}{c}{$\rho=0$} & \multicolumn{3}{c}{$\rho=0.5$} \\
\cmidrule(r){3-5} \cmidrule(lr){6-8} \cmidrule(lr){9-11} \cmidrule(l){12-14}
& & $\nu=50$ & $\nu=100$ & $\nu=200$ & $\nu=50$ & $\nu=100$ & $\nu=200$ & $\nu=50$ & $\nu=100$ & $\nu=200$ & $\nu=50$ & $\nu=100$ & $\nu=200$ \\
\midrule
\multirow{6}{*}{100} 
 & accuracy & 0.744 (0.014) & 0.747 (0.009) & 0.749 (0.007) & 0.729 (0.013) & 0.731 (0.010) & 0.733 (0.008) & 0.695 (0.015) & 0.700 (0.011) & 0.702 (0.007) & 0.668 (0.016) & 0.675 (0.011) & 0.679 (0.009) \\
 & sensitivity & 0.799 (0.022) & 0.808 (0.017) & 0.813 (0.012) & 0.810 (0.023) & 0.814 (0.017) & 0.819 (0.013) & 0.755 (0.023) & 0.762 (0.017) & 0.764 (0.013) & 0.748 (0.026) & 0.758 (0.020) & 0.764 (0.018)\\
 & specificity & 0.674 (0.027) & 0.672 (0.021) & 0.670 (0.014)& 0.629 (0.033) & 0.628 (0.022) & 0.628 (0.020) & 0.619 (0.028) & 0.622 (0.021) & 0.624 (0.016) & 0.568 (0.040) & 0.571 (0.033) & 0.573 (0.032)\\
 & precision & 0.754 (0.019) & 0.754 (0.013) & 0.754 (0.010) & 0.730 (0.018) & 0.729 (0.013) & 0.730 (0.010) & 0.712 (0.019) & 0.715 (0.014) & 0.717 (0.009) & 0.682 (0.018) & 0.685 (0.013) & 0.688 (0.010)\\
 & F1 & 0.776 (0.015) & 0.780 (0.010) & 0.782 (0.008) & 0.768 (0.014) & 0.769 (0.011) & 0.772 (0.009) & 0.733 (0.017) & 0.738 (0.012) & 0.740 (0.009) & 0.713 (0.017) & 0.720 (0.012) & 0.724 (0.011)\\
 & AUC & 0.820 (0.013) & 0.824 (0.009) & 0.826 (0.006) & 0.797 (0.015) & 0.800 (0.011) & 0.803 (0.008) & 0.756 (0.016) & 0.763 (0.011) & 0.765 (0.007) & 0.715 (0.019) & 0.722 (0.014) & 0.726 (0.011)\\
\midrule
\multirow{6}{*}{300} 
 & accuracy & 0.746 (0.008) & 0.748 (0.006) & 0.750 (0.004) & 0.731 (0.008) & 0.732 (0.006) & 0.734 (0.005) & 0.694 (0.009) & 0.699 (0.006) & 0.702 (0.004) & 0.668 (0.009) & 0.675 (0.007) & 0.678 (0.005) \\
 & sensitivity & 0.809 (0.013)& 0.814 (0.010) & 0.818 (0.007) & 0.814 (0.013) & 0.819 (0.010) & 0.822 (0.008) & 0.755 (0.013) & 0.762 (0.010) & 0.765 (0.008) & 0.746 (0.016) & 0.757 (0.013) & 0.763 (0.011) \\
 & specificity & 0.668 (0.016) & 0.667 (0.012) & 0.667 (0.009) & 0.629 (0.018) & 0.626 (0.015) & 0.626 (0.013) & 0.619 (0.016) & 0.621 (0.012) & 0.624 (0.009) & 0.572 (0.023) & 0.574 (0.020) & 0.574 (0.019) \\
 & precision & 0.752 (0.010) & 0.752 (0.008) & 0.753 (0.006) & 0.728 (0.011) & 0.729 (0.008) & 0.730 (0.006) & 0.711 (0.011) & 0.714 (0.008) & 0.716 (0.006) & 0.680 (0.011) & 0.685 (0.008) & 0.688 (0.006) \\
 & F1 & 0.780 (0.008) & 0.782 (0.006) & 0.784 (0.005) & 0.768 (0.009) & 0.771 (0.006) & 0.773 (0.006) & 0.733 (0.009) & 0.737 (0.007) & 0.740 (0.005) & 0.712 (0.010) & 0.719 (0.008) & 0.723 (0.006) \\
 & AUC & 0.823 (0.008) & 0.825 (0.005) & 0.828 (0.004) & 0.800 (0.008) & 0.802 (0.006) & 0.804 (0.005) & 0.756 (0.009) & 0.762 (0.006) & 0.766 (0.004) & 0.715 (0.010) & 0.722 (0.008) & 0.726 (0.006) \\
\midrule
\multirow{6}{*}{500} 
 & accuracy & 0.747 (0.006) & 0.749 (0.005) & 0.751 (0.003) & 0.731 (0.006) & 0.733 (0.004) & 0.734 (0.003) & 0.694 (0.007) & 0.699 (0.005) & 0.701 (0.003) & 0.669 (0.007) & 0.674 (0.005) & 0.678 (0.004)\\
 & sensitivity & 0.812 (0.011) & 0.816 (0.007) & 0.819 (0.005) & 0.816 (0.011) & 0.820 (0.007) & 0.824 (0.006) & 0.755 (0.010) & 0.763 (0.008) & 0.765 (0.006) & 0.747 (0.012) & 0.757 (0.009) & 0.762 (0.008)\\
 & specificity & 0.666 (0.013) & 0.666 (0.010) & 0.666 (0.007) & 0.628 (0.015) & 0.625 (0.011) & 0.624 (0.010) & 0.618 (0.012) & 0.621 (0.009) & 0.622 (0.007) & 0.573 (0.018) & 0.572 (0.017) & 0.574 (0.014)\\
 & precision & 0.752 (0.008) & 0.752 (0.006) & 0.753 (0.004) & 0.729 (0.009) & 0.729 (0.006) & 0.729 (0.004) & 0.711 (0.009) & 0.715 (0.006) & 0.716 (0.004) & 0.682 (0.009) & 0.685 (0.007) & 0.687 (0.004)\\
 & F1 & 0.780 (0.007) & 0.783 (0.005) & 0.784 (0.003) & 0.770 (0.007) & 0.772 (0.005) & 0.773 (0.004) & 0.733 (0.007) & 0.738 (0.005) & 0.739 (0.004) & 0.713 (0.008) & 0.719 (0.005) & 0.723 (0.004) \\
 & AUC & 0.823 (0.006) & 0.826 (0.004) & 0.828 (0.003) & 0.801 (0.007) & 0.802 (0.005) & 0.805 (0.004) & 0.756 (0.007) & 0.762 (0.005) & 0.765 (0.003) & 0.715 (0.008) & 0.722 (0.007) & 0.726 (0.005)\\
\bottomrule
\bottomrule
\end{tabular}
}
\end{sidewaystable}

\subsection{Comparing stochastic optimization methods}
\label{s:simstooptmethods}

We compare the SSAMSGrad with two alternative implementations of the GPLM, the stratified sampling Adam (SSAdam), and stratified sampling root mean square propagation \citep[SSRMSProp,][]{hinton2012neural} in terms of predictive power. The latter two algorithms, as well as stratified sampling SGD (SSSGD), are described in Appendix C of the Supplementary Material. Simulated data are generated according to the following mechanism:

\begin{itemize}
\item Provider count $m = 100$, or 300;
\item Provider-specific subject counts $\{n_i: i = 1, \ldots, m\}$ are drawn from a Poisson distribution with the mean $\nu$ equal to 50 or 100, and are truncated to be at least 20;
\item Provider effects $\bgamma$ are sampled from a Gaussian distribution $\mathcal{N}(\mu, \sigma^2)$ with $\mu = \log(4/11)$ and $\sigma = 0.4$, and are fixed throughout all simulated data sets;
\item Subject-specific covariates $\bZ_{ij}$ are generated according to formula \eqref{eq:samplecovar} with $\rho = 0.5$;
\item A function of linear associations is specified according to model \eqref{eq:linear};
\item A function of linear and nonlinear associations is specified according to model \eqref{eq:nonlinear};
\item The outcome $Y_{ij}$ is sampled from $\mathrm{Bernoulli}(\expit\{\gamma_i + g_k^*(\bZ_{ij})\})$, where $k = 1, 2$; and
\item In each scenario, 500 simulated data sets are generated.
\end{itemize}

We fit the three stratified sampling-based algorithms of the fixed-effect GPLM to each simulated data set to gauge predictive power. In addition, we calculate the ratio of the time to convergence for SSAdam (SSRMSProp, respectively) to the time to convergence for SSAMSGrad, also known as the speedup of the SSAMSGrad relative to SSAdam (SSRMSProp, respectively). All FNNs include an input layer of 3 nodes, two hidden layers of 32 and 16 nodes, respectively, and an output layer of 1 node. The corresponding activation functions are ReLU, ReLU, and identity, respectively.

\begin{table}[htbp]
\caption{Performance of the SSAMSGrad, SSAdam, and SSRMSProp under generalized partially linear model (GPLM). For each predictive metric, the mean and standard deviation (in parentheses) of each metric are derived from 500 simulated data sets. The average speedup is calculated as follows: (1) for each algorithm, the average time to convergence is measured across five model fits with the same simulated data; (2) speedups of the SSAMSGrad relative to the SSAdam and SSRMSProp are calculated respectively for each simulated data; (3) speedups are averaged across 50 simulated data sets. \label{tab:stooptmethods}}
\centering
\resizebox{\columnwidth}{!}{%
\begin{tabular}{cccccccccccccccc}
\toprule
\toprule
\multicolumn{8}{c}{Panel A: $g^*_1(\bZ_{ij})$ in \eqref{eq:linear}} \\
\midrule
\multirow{3}{*}{$m$} & \multirow{3}{*}{metric} & \multicolumn{2}{c}{SSAMSGrad} & \multicolumn{2}{c}{SSAdam} & \multicolumn{2}{c}{SSRMSProp} \\
\cmidrule(r){3-4} \cmidrule(lr){5-6} \cmidrule(l){7-8}
& & $\nu=50$ & $\nu=100$ & $\nu=50$ & $\nu=100$ & $\nu=50$ & $\nu=100$ \\
\midrule
\multirow{7}{*}{100} 
 & accuracy & 0.744 (0.015) & 0.746 (0.011)  & 0.745 (0.015) & 0.746 (0.011) & 0.743 (0.015) & 0.746 (0.012) \\
 & sensitivity & 0.880 (0.020) & 0.885 (0.015) & 0.880 (0.020) & 0.883 (0.015) & 0.881 (0.021) & 0.884 (0.016)\\
 & specificity & 0.449 (0.042) & 0.445 (0.032) & 0.452 (0.044) & 0.449 (0.031) & 0.444 (0.047) & 0.445 (0.034) \\
 & precision & 0.777 (0.017) & 0.776 (0.012) & 0.777 (0.017) & 0.777 (0.012) & 0.775 (0.017) & 0.775 (0.013) \\
 & F1 & 0.825 (0.013) & 0.827 (0.010) & 0.825 (0.013) & 0.826 (0.010) & 0.824 (0.013) & 0.826 (0.011) \\
 & AUC & 0.773 (0.016) & 0.776 (0.011) & 0.773 (0.016) & 0.776 (0.011) & 0.772 (0.016) & 0.775 (0.012) \\
 & speedup & 1 & 1 & 2.416 & 2.905 & 1.015 & 1.147\\ 
\midrule
\multirow{7}{*}{300} 
 & accuracy & 0.746 (0.009) & 0.748 (0.006) & 0.746 (0.009) & 0.747 (0.006) & 0.746 (0.009) & 0.747 (0.007) \\
 & sensitivity & 0.886 (0.012) & 0.890 (0.008) & 0.884 (0.011) & 0.888 (0.009) & 0.885 (0.012) & 0.888 (0.009) \\
 & specificity & 0.442 (0.026) & 0.438 (0.018) & 0.447 (0.023) & 0.441 (0.020) & 0.444 (0.025) & 0.441 (0.020) \\
 & precision & 0.775 (0.010) & 0.775 (0.007) & 0.776 (0.010) & 0.775 (0.007) & 0.775 (0.009) & 0.775 (0.007) \\
 & F1 & 0.827 (0.008) & 0.829 (0.006) & 0.826 (0.008) & 0.827 (0.006) & 0.826 (0.007) & 0.828 (0.006) \\
 & AUC & 0.776 (0.009) & 0.777 (0.007) & 0.775 (0.009) & 0.776 (0.007) & 0.774 (0.009) & 0.778 (0.007) \\
  & speedup & 1 & 1 & 2.959 & 3.138 & 1.109  & 1.033 \\ 
\toprule
\multicolumn{8}{c}{Panel B: $g^*_2(\bZ_{ij})$ in \eqref{eq:nonlinear}} \\
\midrule
\multirow{3}{*}{$m$} & \multirow{3}{*}{metric} & \multicolumn{2}{c}{SSAMSGrad} & \multicolumn{2}{c}{SSAdam} & \multicolumn{2}{c}{SSRMSProp}  \\
\cmidrule(r){3-4} \cmidrule(lr){5-6} \cmidrule(lr){7-8}
& & $\nu=50$ & $\nu=100$ & $\nu=50$ & $\nu=100$ & $\nu=50$ & $\nu=100$ \\
\midrule
\multirow{7}{*}{100} 
 & accuracy & 0.728 (0.014) & 0.731 (0.010) & 0.727 (0.014) & 0.731 (0.011) & 0.727 (0.015) & 0.730 (0.011)  \\
 & sensitivity & 0.807 (0.025) & 0.813 (0.017) & 0.804 (0.026) & 0.812 (0.018) & 0.806 (0.024) & 0.812 (0.017) \\
 & specificity & 0.630 (0.033) & 0.630 (0.024) & 0.633 (0.033) & 0.631 (0.027) & 0.629 (0.033) & 0.630 (0.025) \\
 & precision & 0.728 (0.019) & 0.731 (0.014) & 0.729 (0.019) & 0.730 (0.014) & 0.728 (0.020) & 0.729 (0.014) \\
 & F1 & 0.765 (0.016) & 0.770 (0.012) & 0.764 (0.016) & 0.768 (0.012) & 0.765 (0.016)  & 0.768 (0.012) \\
 & AUC & 0.797 (0.015) & 0.801 (0.011) & 0.797 (0.015) & 0.800 (0.011) & 0.796 (0.015) & 0.800 (0.011) \\
 & speedup & 1 & 1 & 3.189 & 3.016 & 1.135 & 1.180 \\ 
\midrule
\multirow{7}{*}{300} 
 & accuracy & 0.730 (0.008) & 0.733 (0.006) & 0.730 (0.008) & 0.732 (0.006)  & 0.729 (0.009) & 0.733 (0.006) \\
 & sensitivity & 0.814 (0.014) & 0.817 (0.010) & 0.812 (0.014) & 0.817 (0.011) & 0.813 (0.013) & 0.817 (0.011) \\
 & specificity & 0.628 (0.019) & 0.628 (0.015) & 0.629 (0.019) & 0.628 (0.015) & 0.627 (0.019) & 0.628 (0.015) \\
 & precision & 0.729 (0.011) & 0.729 (0.008) & 0.729 (0.011) & 0.729 (0.008) & 0.727 (0.012) & 0.730 (0.008) \\
 & F1 & 0.769 (0.009) & 0.771 (0.006) & 0.768 (0.009) & 0.771 (0.007) & 0.768 (0.009) & 0.771 (0.007)\\
 & AUC & 0.799 (0.008) & 0.803 (0.006) & 0.799 (0.009) & 0.802 (0.006) & 0.798 (0.009) & 0.802 (0.006) \\
 & speedup & 1 & 1 & 3.008 & 2.985 & 1.134 & 0.847 \\ 
\bottomrule
\bottomrule
\end{tabular}
}
\end{table}

\Cref{tab:stooptmethods} compares SSAMSGrad with SSAdam and SSRMSProp in terms of accuracy, sensitivity, specificity, precision, F1, AUC, and speedup. For a given algorithm, an increase in $m$ or $\nu$ generally leads to higher values in all predictive metrics but sensitivity. Across all settings, the three algorithms share nearly identical values of all the predictive metrics. However, the SSAMSGrad is at least two times as fast as the SSAdam and is generally faster than the SSRMSProp. These findings, consistent with the early work \citep{reddi2018amsgrad}, confirm the outstanding performance of the SSAMSGrad among the three stratified sampling-based algorithms. SSSGD is not considered here since it is considerably slower than the other three methods (details in Table 1 of the Supplementary Material). For comparison, results from the simple sampling-based algorithms are included in Table 2 of the Supplementary Material, where AMSGrad again stands out as the fastest approach.

\subsection{Comparing sampling schemes}

To study the advantage of stratified sampling in stochastic optimization, we compare the SSAMSGrad, SSAdam, and SSRMSProp with their simple sampling counterparts. The data-generating mechanism is the same as in \Cref{s:simstooptmethods}. We fit six stratified and simple sampling algorithms of the fixed-effect GPLM to each simulated data set. All FNNs include an input layer of 3 nodes, two hidden layers of 32 and 16 nodes, respectively, and an output layer of 1 node. The corresponding activation functions are ReLU, ReLU, and identity, respectively. For each model fit, we consider two performance metrics. The first metric is the variance of gradient components derived from an iteration of an algorithm, and the second metric is the speedup of the stratified sampling algorithm relative to its simple sampling counterpart. Following the notation in \Cref{s:stoopt}, let $\cT_i = \{(i', j') \in \cT: i' = i\}$ and $\cT_{i(s)} = \{(i', j') \in \cT_{(s)}: i' = i\}$, where $i$ indicates provider $i$ and $s$ indicates the $s$th iteration of an algorithm. At iteration $s$, the variance of gradient components under stratified sampling is
\[
\frac{1}{|\cT|^2} \sum_{i=1}^m \frac{|\cT_i|}{|\cT_{i(s)}|}\sum_{(i, j) \in \cT_{i}} \left\|\dot\ell_{ij}(\btheta_{(s-1)}) - \frac{1}{|\cT_i|}\sum_{(i, j') \in \cT_i} \dot\ell_{ij'}(\btheta_{(s-1)})\right\|^2,
\]
while the variance of gradient components under simple sampling is
\[
\frac{1}{|\cT|\cdot|\cT_{(s)}|} \sum_{(i, j) \in \cT_{(s)}} \left\|\dot\ell_{ij}(\btheta_{(s-1)}) - \frac{1}{|\cT|}\sum_{(i', j') \in \cT} \dot\ell_{i'j'}(\btheta_{(s-1)})\right\|^2.
\]

\begin{sidewaystable}[htbp]
\caption{Variance of gradient components and speedup of stratified and simple random sampling-based AMSGrad, Adam, and RMSProp under the generalized partially linear model. The mean and standard deviation (in parentheses) of each variance metric (with respect to the weights $\bw$, biases $\bb$, and provider effects $\bgamma$) are derived from 500 simulated data sets. The average speedup is calculated as follows: (1) for each algorithm, the average time to convergence under each sampling scheme is measured across five model fits with the same simulated data; (2) speedups of stratified sampling relative to simple sampling are calculated for each simulated data; (3) speedups are averaged across 50 simulated data sets. \label{tab:sampling}}
\centering
\resizebox{\columnwidth}{!}{%
\begin{tabular}{cccccccccccccccc}
\toprule
\toprule
\multicolumn{14}{c}{Panel A: $g^*_1(\bZ_{ij})$ in \eqref{eq:linear}} \\
\midrule
\multirow{3}{*}{$m$} & \multirow{3}{*}{metric} & \multicolumn{4}{c}{AMSGrad} & \multicolumn{4}{c}{Adam} & \multicolumn{4}{c}{RMSProp} \\
\cmidrule(r){3-6} \cmidrule(lr){7-10} \cmidrule(l){11-14}
& & \multicolumn{2}{c}{stratified} & \multicolumn{2}{c}{simple} & \multicolumn{2}{c}{stratified} & \multicolumn{2}{c}{simple} & \multicolumn{2}{c}{stratified} & \multicolumn{2}{c}{simple} \\
\cmidrule(r){3-4} \cmidrule(lr){5-6} \cmidrule(lr){7-8} \cmidrule(lr){9-10} \cmidrule(lr){11-12} \cmidrule(l){13-14}
& & $\nu=50$ & $\nu=100$ & $\nu=50$ & $\nu=100$ & $\nu=50$ & $\nu=100$ & $\nu=50$ & $\nu=100$ & $\nu=50$ & $\nu=100$ & $\nu=50$ & $\nu=100$ \\
\midrule
\multirow{7}{*}{100} 
 & \multirow{2}{*}{variance ($\bw$)} & $9.09 \times 10^{-7}$ & $4.04 \times 10^{-7}$ & $9.35 \times 10^{-7}$ & $4.16 \times 10^{-7}$ & $9.04 \times 10^{-7} $  & $4.03 \times 10^{-7} $ & $9.29 \times 10^{-7}$ & $4.15 \times 10^{-7}$ & $9.09 \times 10^{-7}$ & $4.01 \times 10^{-7}$ & $9.37 \times 10^{-7}$ & $4.12 \times 10^{-7}$ \\
 & & $(1.05 \times 10^{-7})$ & $(4.47 \times 10^{-8})$ & $(1.08 \times 10^{-7})$ & $(4.74 \times 10^{-8})$ & $(1.12 \times 10^{-7})$ & $(4.45 \times 10^{-8})$ & $(1.17 \times 10^{-7})$ & $(4.73 \times 10^{-8})$ & $(1.07 \times 10^{-7})$ & $(4.31 \times 10^{-8})$ & $(1.13 \times 10^{-7})$ & $(4.57 \times 10^{-8})$ \\
 & \multirow{2}{*}{variance ($\bb$)} & $1.19 \times 10^{-3}$ & $5.31 \times 10^{-4}$ & $1.27 \times 10^{-3} $ & $5.67 \times 10^{-4}$ & $1.19 \times 10^{-3}$ & $5.30 \times 10^{-4}$ & $1.27 \times 10^{-3} $ & $5.67 \times 10^{-4}$ & $1.19 \times 10^{-3}$ & $5.31 \times 10^{-4}$ & $1.27 \times 10^{-3} $ & $5.67 \times 10^{-4}$ \\
 & & $(4.00 \times 10^{-5}) $ & $(1.31 \times 10^{-5})$ & $(3.90 \times 10^{-5}) $ & $(1.40 \times 10^{-5})$ & $(4.63 \times 10^{-5}) $ & $(1.36 \times 10^{-5})$ & $(4.55 \times 10^{-5}) $ & $(1.42 \times 10^{-5})$ & $(4.46 \times 10^{-5}) $ & $(1.29 \times 10^{-5})$ & $(4.18 \times 10^{-5}) $ & $(1.34 \times 10^{-5})$\\
 & \multirow{2}{*}{variance ($\bgamma$)} & $ 1.05 \times 10^{-3}$ & $4.66 \times 10^{-4}$ & $1.13 \times 10^{-3} $ & $5.02 \times 10^{-4}$ & $1.04 \times 10^{-3}$ & $4.67 \times 10^{-4}$ & $1.12 \times 10^{-3}$ & $5.02 \times 10^{-4}$ & $1.05 \times 10^{-3}$ & $4.67 \times 10^{-4}$ & $1.13 \times 10^{-3}$ & $5.02 \times 10^{-4}$ \\
 & & $(3.67 \times 10^{-5}) $ & $(1.22 \times 10^{-5})$ & $(3.59 \times 10^{-5}) $ & $(1.31 \times 10^{-5})$ & $(4.17 \times 10^{-5})$ & $(1.27 \times 10^{-5})$ & $(4.11 \times 10^{-5})$ & $(1.33 \times 10^{-5})$ & $(4.11 \times 10^{-5})$ & $(1.20 \times 10^{-5})$ & $(3.84 \times 10^{-5})$ & $(1.24 \times 10^{-5})$\\
 & speedup & 1 & 1 & 1.174 & 1.378 & 1 & 1 & 1.093 & 1.150 & 1 & 1 & 1.258 & 1.161 \\
\midrule
\multirow{7}{*}{300} 
 & \multirow{2}{*}{variance ($\bw$)} & $3.00 \times 10^{-7}$ & $1.35 \times 10^{-7}$ & $3.09 \times 10^{-7}$ & $1.39 \times 10^{-7}$ & $3.02 \times 10^{-7} $  & $1.35 \times 10^{-7} $ & $3.11 \times 10^{-7}$ & $1.39 \times 10^{-7}$ & $3.02 \times 10^{-7}$ & $1.34 \times 10^{-7}$ & $3.11 \times 10^{-7}$ & $1.39 \times 10^{-7}$ \\
 & & $(2.01 \times 10^{-8})$ & $(8.36 \times 10^{-9})$ & $(2.08 \times 10^{-8})$ & $(8.83 \times 10^{-9})$ & $(2.17 \times 10^{-8})$ & $(8.58 \times 10^{-9})$ & $(2.25 \times 10^{-8})$ & $(9.04 \times 10^{-9})$ & $(2.02 \times 10^{-8})$ & $(8.57 \times 10^{-9})$ & $(2.11 \times 10^{-8})$ & $(9.05 \times 10^{-9})$ \\
 & \multirow{2}{*}{variance ($\bb$)} & $3.96 \times 10^{-4}$ & $1.77 \times 10^{-4}$ & $4.23 \times 10^{-4}$ & $1.89 \times 10^{-4}$ & $3.97 \times 10^{-4}$ & $1.77 \times 10^{-4}$ & $4.24 \times 10^{-4}$ & $1.89 \times 10^{-4}$ & $3.97 \times 10^{-4}$ & $1.77 \times 10^{-4}$ & $4.24 \times 10^{-4}$ & $1.89 \times 10^{-4}$ \\
 & & $(8.49 \times 10^{-6})$ & $(2.58 \times 10^{-6})$ & $(8.30 \times 10^{-6})$ & $(2.71 \times 10^{-6})$ & $(8.26 \times 10^{-6})$ & $(2.42 \times 10^{-6})$ & $(7.89 \times 10^{-6})$ & $(2.52 \times 10^{-6})$ & $(8.17 \times 10^{-6})$ & $(2.45 \times 10^{-6})$ & $(7.93 \times 10^{-6})$ & $(2.61 \times 10^{-6})$ \\
 & \multirow{2}{*}{variance ($\bgamma$)} & $3.48 \times 10^{-4}$ & $1.56 \times 10^{-4}$ & $3.74 \times 10^{-4}$ & $1.67 \times 10^{-4}$ & $3.49 \times 10^{-4}$ & $1.55 \times 10^{-4}$ & $3.75 \times 10^{-4}$ & $1.67 \times 10^{-4}$ & $3.49 \times 10^{-4}$ & $1.56 \times 10^{-4}$ & $3.75 \times 10^{-4}$ & $1.67 \times 10^{-4}$ \\
 & & $(7.61 \times 10^{-6})$ & $(2.46 \times 10^{-6})$ & $(7.51 \times 10^{-6})$ & $(2.57 \times 10^{-6})$ & $(7.47 \times 10^{-6})$ & $(2.29 \times 10^{-6})$ & $(7.13 \times 10^{-6})$ & $(2.39 \times 10^{-6})$ & $(7.41 \times 10^{-6})$ & $(2.28 \times 10^{-6})$ & $(7.14 \times 10^{-6})$ & $(2.42 \times 10^{-6})$ \\ 
 & speedup & 1 & 1 & 1.205 & 1.151 & 1 & 1 & 1.121 & 1.061 & 1 & 1 & 1.138 & 1.071 \\
\toprule
\multicolumn{14}{c}{Panel B: $g^*_2(\bZ_{ij})$ in \eqref{eq:nonlinear}} \\
\midrule
\multirow{3}{*}{$m$} & \multirow{3}{*}{metric} & \multicolumn{4}{c}{AMSGrad} & \multicolumn{4}{c}{Adam} & \multicolumn{4}{c}{RMSProp}  \\
\cmidrule(r){3-6} \cmidrule(lr){7-10} \cmidrule(l){11-14}
& & \multicolumn{2}{c}{stratified} & \multicolumn{2}{c}{simple} & \multicolumn{2}{c}{stratified} & \multicolumn{2}{c}{simple} & \multicolumn{2}{c}{stratified} & \multicolumn{2}{c}{simple} \\
\cmidrule(r){3-4} \cmidrule(lr){5-6} \cmidrule(lr){7-8} \cmidrule(lr){9-10} \cmidrule(lr){11-12} \cmidrule(l){13-14}
& & $\nu=50$ & $\nu=100$ & $\nu=50$ & $\nu=100$ & $\nu=50$ & $\nu=100$ & $\nu=50$ & $\nu=100$ & $\nu=50$ & $\nu=100$ & $\nu=50$ & $\nu=100$ \\
\midrule
\multirow{7}{*}{100} 
 & \multirow{2}{*}{variance ($\bw$)} & $8.20 \times 10^{-7}$ & $3.64 \times 10^{-7}$ & $9.10 \times 10^{-7}$ & $4.04 \times 10^{-7}$ & $8.09 \times 10^{-7} $  & $3.64 \times 10^{-7} $ & $8.99 \times 10^{-7}$ & $4.05 \times 10^{-7}$ & $8.14 \times 10^{-7}$ & $3.65 \times 10^{-7}$ & $9.03 \times 10^{-7}$ & $4.05 \times 10^{-7}$ \\
 & & $(8.16 \times 10^{-8})$ & $(3.47 \times 10^{-8})$ & $(9.95 \times 10^{-8})$ & $(4.37 \times 10^{-8})$ & $(8.07 \times 10^{-8})$ & $(3.51 \times 10^{-8})$ & $(9.82 \times 10^{-8})$ & $(4.37 \times 10^{-8})$ & $(8.14 \times 10^{-8})$ & $(3.40 \times 10^{-8})$ & $(9.92 \times 10^{-8})$ & $(4.20 \times 10^{-8})$ \\
  & \multirow{2}{*}{variance ($\bb$)} & $1.34 \times 10^{-3}$ & $5.99 \times 10^{-4}$ & $1.44 \times 10^{-3}$ & $6.40 \times 10^{-4}$ & $1.34 \times 10^{-3}$ & $6.00 \times 10^{-4}$ & $1.43 \times 10^{-3}$ & $6.41 \times 10^{-4}$ & $1.34 \times 10^{-3}$ & $5.99 \times 10^{-4}$ & $1.43 \times 10^{-3}$ & $6.41 \times 10^{-4}$ \\
 & & $(4.06 \times 10^{-5})$ & $(1.03 \times 10^{-5})$ & $(3.51 \times 10^{-5})$ & $(9.84 \times 10^{-6})$ & $(4.06 \times 10^{-5})$ & $(1.03 \times 10^{-5})$ & $(3.51 \times 10^{-5})$ & $(9.84 \times 10^{-6})$ & $(3.78 \times 10^{-5})$ & $(1.00 \times 10^{-5})$ & $(3.36 \times 10^{-5})$ & $(9.18 \times 10^{-6})$ \\
 & \multirow{2}{*}{variance ($\bgamma$)} & $1.20 \times 10^{-3}$ & $5.35 \times 10^{-4}$ & $1.29 \times 10^{-3}$ & $5.75 \times 10^{-4}$ & $1.20 \times 10^{-3}$  & $5.35 \times 10^{-4}$ & $1.29 \times 10^{-3}$ & $5.76 \times 10^{-4}$ & $1.20 \times 10^{-3}$ & $5.35 \times 10^{-4}$ & $1.29 \times 10^{-3}$ & $5.75 \times 10^{-4}$ \\
 & & $(3.49 \times 10^{-5})$ & $(9.01 \times 10^{-6})$ & $(3.02 \times 10^{-5})$ & $(7.87 \times 10^{-6})$ & $(3.53 \times 10^{-5})$ & $(9.16 \times 10^{-6})$ & $(2.98 \times 10^{-5})$ & $(8.59 \times 10^{-6})$ & $(3.32 \times 10^{-5})$ & $(8.70 \times 10^{-6})$ & $(2.90 \times 10^{-5})$ & $(7.77 \times 10^{-6})$ \\
 & speedup & 1 & 1 & 1.107 & 1.220 & 1 & 1 & 1.223 & 1.125 & 1 & 1 & 1.245 & 1.204 \\
\midrule
\multirow{7}{*}{300} 
 & \multirow{2}{*}{variance ($\bw$)} & $2.71 \times 10^{-7}$ & $1.21 \times 10^{-7}$ & $3.01 \times 10^{-7}$ & $1.35 \times 10^{-7}$ & $2.71 \times 10^{-7}$ & $1.21 \times 10^{-7}$ & $3.01 \times 10^{-7}$ & $1.35 \times 10^{-7}$ & $2.72 \times 10^{-7}$ & $1.22 \times 10^{-7}$ & $3.01 \times 10^{-7}$ & $1.35 \times 10^{-7}$  \\
 & & $(1.63 \times 10^{-8})$ & $(6.62 \times 10^{-9})$ & $(2.02 \times 10^{-8})$ & $(8.38 \times 10^{-9})$ & $(1.63 \times 10^{-8})$ & $(6.27 \times 10^{-9})$ & $(1.97 \times 10^{-8})$ & $(8.02 \times 10^{-9})$ & $(1.56 \times 10^{-8})$ & $(6.56 \times 10^{-9})$ & $(1.92 \times 10^{-8})$ & $(8.40 \times 10^{-9})$ \\
  & \multirow{2}{*}{variance ($\bb$)} & $4.48 \times 10^{-4}$ & $2.00 \times 10^{-4}$ & $4.79 \times 10^{-4}$ & $2.14 \times 10^{-4}$ & $4.48 \times 10^{-4}$ & $2.00 \times 10^{-4}$ & $4.79 \times 10^{-4}$ & $2.14 \times 10^{-4}$ & $4.48 \times 10^{-4}$ & $2.00 \times 10^{-4}$ & $4.79 \times 10^{-4}$ & $2.14 \times 10^{-4}$ \\
 & & $(7.10 \times 10^{-6})$ & $(1.90 \times 10^{-6})$ & $(6.57 \times 10^{-6})$ & $(1.80 \times 10^{-6})$ & $(7.96 \times 10^{-6})$ & $(1.96 \times 10^{-6})$ & $(7.11 \times 10^{-6})$ & $(1.78 \times 10^{-6})$ & $(7.29 \times 10^{-6})$ & $(2.05 \times 10^{-6})$ & $(6.40 \times 10^{-6})$ & $(1.77 \times 10^{-6})$ \\
 & \multirow{2}{*}{variance ($\bgamma$)} & $4.00 \times 10^{-4}$ & $1.78 \times 10^{-4}$ & $4.29 \times 10^{-4}$ & $1.92 \times 10^{-4}$ & $4.00 \times 10^{-4}$ & $1.78 \times 10^{-4}$ & $4.30 \times 10^{-4}$ & $1.92 \times 10^{-4}$ & $4.00 \times 10^{-4}$ & $1.78 \times 10^{-4}$ & $4.30 \times 10^{-4}$ & $1.92 \times 10^{-4}$ \\
 & & $(6.94 \times 10^{-6})$ & $(1.77 \times 10^{-6})$ & $(6.16 \times 10^{-6})$ & $(1.53 \times 10^{-6})$ & $(6.20 \times 10^{-6})$ & $(1.66 \times 10^{-6})$ & $(5.64 \times 10^{-6})$ & $(1.53 \times 10^{-6})$ & $(6.47 \times 10^{-6})$ & $(1.82 \times 10^{-6})$ & $(5.64 \times 10^{-6})$ & $(1.50 \times 10^{-6})$ \\
 & speedup & 1 & 1 & 1.200 & 1.210 & 1 & 1 & 1.110 & 1.167 & 1 & 1 & 1.248 & 1.350 \\
\bottomrule
\bottomrule
\end{tabular}
}
\end{sidewaystable}

\Cref{tab:sampling} shows the average variance of gradient components and speedup for different values of $m$ and $\nu$, optimization methods, sampling schemes, and underlying models. The variance of gradient components is calculated separately for the weights $\bw$, biases $\bb$, and provider effects $\bgamma$. For all three algorithms, stratified sampling is always associated with lower variance metrics than simple sampling. The average speedup for all algorithms is greater than one in all settings, indicating that stratified sampling leads to accelerated convergence.

\subsection{Comparing tests for outlier identification}

Lastly, we compare the exact test with score and Wald tests in detecting outlying health care providers with unusual performance. We consider the following data-generating mechanism:

\begin{itemize}
\item Provide count $m = 100$;
\item Provider-specific subject counts $\{n_i: i = 2, \ldots, m\}$ are drawn from $\mathrm{Poisson}(100)$, and are truncated to be at least 20, while $n_1$ is set to 20, 50, or 100;
\item Provider effects $\{\gamma_i: i = 2, \ldots, m\}$ are sampled from a Gaussian distribution $\mathcal{N}(\mu, \sigma^2)$ with $\mu = \log(4/11)$ and $\sigma = 0.4$; the effect $\gamma_1 = \mu$ for settings where the type I error rate is concerned, and $\gamma_1 = \mu + \Delta \sigma$ with $\Delta$ being an integer varying from $-4$ to 4 when the power is concerned; all effects are fixed throughout all simulated data sets in a single scenario;
\item Subject-specific covariates $\bZ_{ij}$ are generated according to \eqref{eq:samplecovar}, with $\rho$ varying from 0 to 0.9 by 0.1 for settings on type I error rate and $\rho = 0.5$ for settings on power;
\item A function of linear and nonlinear associations is specified as
\[
g_2^*(\bZ_{ij}) = Z_{ij1} + 0.5Z_{ij2} - Z_{ij3} + 0.01 Z_{ij1}Z_{ij2} + 0.01 Z^2_{ij2} + 0.1 \cos(Z_{ij1}) \sin(Z_{ij3});
\]
\item The outcome $Y_{ij}$ is sampled from $\mathrm{Bernoulli}(\expit\{\gamma_i + g_2^*(\bZ_{ij})\})$; and
\item In each scenario, 1,000 simulated data sets are generated.
\end{itemize}

We fit the fixed-effect GPLM to each simulated data set with the GPLM implemented as SSAMSGrad in \Cref{alg:SSAMSGrad}. All FNNs include an input layer of 3 nodes, two hidden layers of 32 and 16 nodes, respectively, and an output layer of 1 node. As before, the corresponding activation functions are ReLU, ReLU, and identity, respectively. For each model fit, we conduct exact, score, and Wald tests regarding the null hypothesis $H_{0,1}: \gamma_1 = \mathrm{median}(\bgamma)$, with a significance level of 0.05. The exact test procedure follows \Cref{subs:exacttest}; the score test statistic is given by
\[
\frac{\sum_{j=1}^{n_1} [Y_{1j} - \expit\{\mathrm{median}(\hat\bgamma) + g_2(\bZ_{1j}; \hat\bw, \hat\bb)\}]}{\sqrt{\sum_{j=1}^{n_1} \expit\{\mathrm{median}(\hat\bgamma) + g_2(\bZ_{1j}; \hat\bw, \hat\bb)\}[1-\expit\{\mathrm{median}(\hat\bgamma) + g_2(\bZ_{1j}; \hat\bw, \hat\bb)\}]}};
\]
the Wald test statistic is given by
\[
\{\hat\gamma_1 - \mathrm{median}(\hat\bgamma)\}
\sqrt{\sum_{j=1}^{n_1} \expit\{\hat\gamma_1 + g_2(\bZ_{1j}; \hat\bw, \hat\bb)\}[1-\expit\{\hat\gamma_1 + g_2(\bZ_{1j}; \hat\bw, \hat\bb)\}]}.
\]
The test statistics are compared with the standard Gaussian distribution to determine whether $H_{0,1}$ should be rejected. Two-sided type I error rate and power are used as performance metrics.

\begin{figure}[hbtp]
\centering
\includegraphics[width=0.75\textwidth]{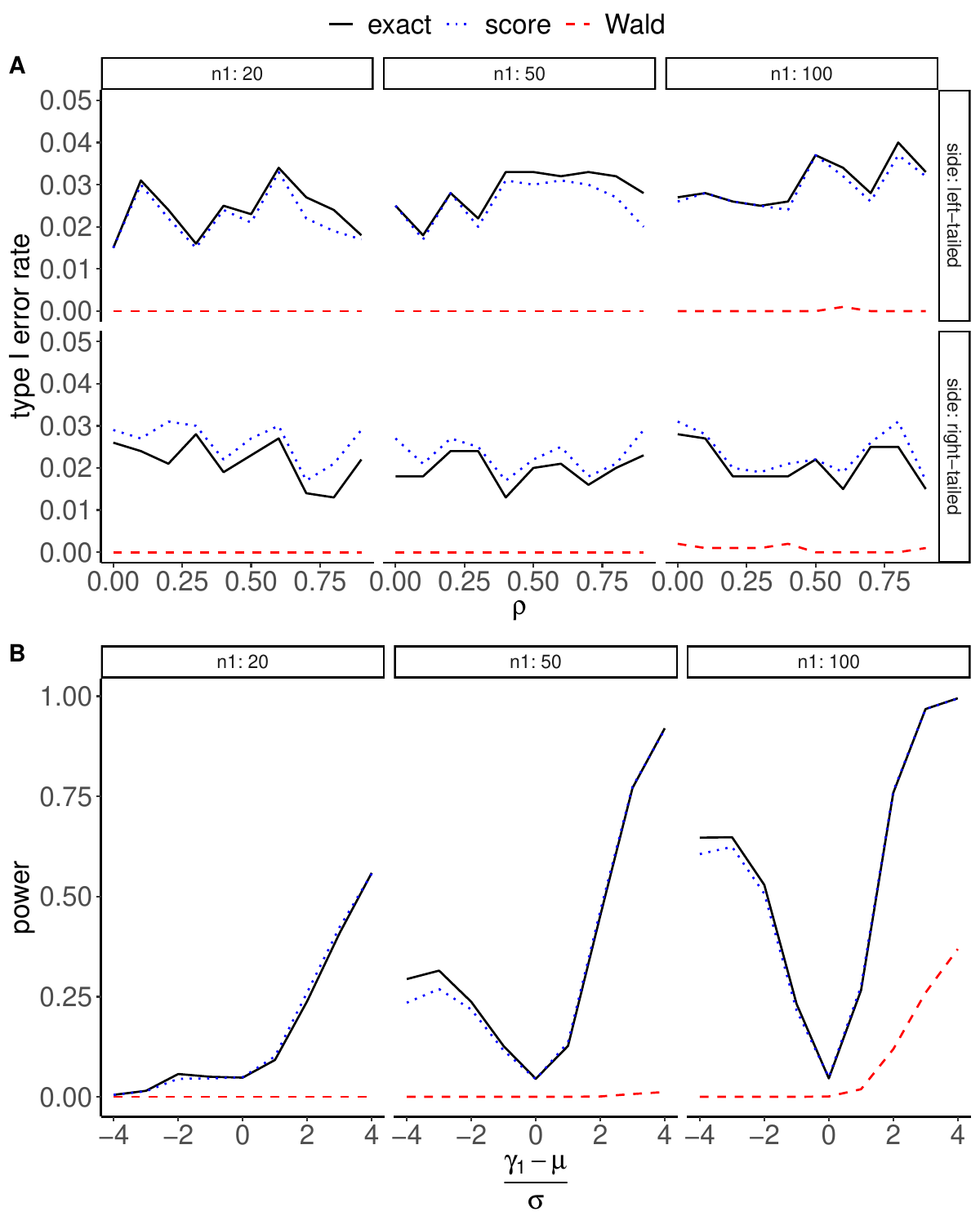}
\caption{Type I error rate and power of the exact, score, and Wald tests. All values are calculated based on 1,000 simulated data sets with a significance level of 0.05.}
\label{fig:errpower}
\end{figure}

Shown in Panel A of \Cref{fig:errpower}, the exact test has a slightly higher left-tailed type I error rate than the score test, whereas the Wald test has an unusually low left-tailed type I error rate close to 0. As the correlation $\rho$ varies from 0 to 0.9, the left-tailed error rate fluctuates around the nominal level of 0.025 when $n_1 = 20$, but tends to exceed the nominal level when $n_1 = 50$ and 100. The exact test tends to have a lower level of right-tailed type I error rate than the score test across different levels of $n_1$ and $\rho$, indicating the right-tailed conservativeness of the exact test. Again, both tests have a reasonably higher right-tailed error rate than the Wald test. Panel B of \Cref{fig:errpower} displays the power as a function of the relative deviation $(\gamma_1 - \mu)/\sigma$ of the true provider effect $\gamma_1$. As expected, a higher relative deviation in magnitude is associated with a higher power, with a positive deviation leading to a higher increase in power than a negative deviation of the same magnitude. This is particularly advantageous when it is of primary interest to test whether $\gamma_1$ is significantly higher than the median provider effect. The exact test has a slightly higher power than the score test when the deviation is negative, and has a similar power to the score test when the deviation is positive. In contrast, the Wald test has a much lower power than the other two tests.

\section{Application to Medicare claims data}
\label{s:app}
\subsection{Medicare inpatient claims for ESRD beneficiaries on kidney dialysis}
We apply the neural network profiling methodology to Medicare inpatient claims for ESRD beneficiaries undergoing kidney dialysis in the year 2020. These claims were sourced from the United States Renal Data System \citep[USRDS,][]{usrds2022adr}. The outcome of interest was all-cause unplanned hospital readmission within 30 days of discharge. Planned readmissions, not deemed unplanned readmissions, were ruled out based on a list of diagnosis codes of the International Classification of Diseases, 10th Revision (ICD-10), available in Appendix E of the Supplementary Material). In addition to the date of discharge, outcome, and USRDS-assigned Medicare-certified dialysis facility identifier, the data set consisted of patient demographic (sex, age at first ESRD service, race, and ethnicity), physical (body mass index and functional status), social (substance/alcohol/tobacco use and employment status), and clinical characteristics (length of hospital stay, time since ESRD diagnosis, and dialysis mode), cause of ESRD (diabetes, hypertension, primary glomerulonephritis, or other), and prevalent comorbidities (in-hospital COVID-19, heart failure, coronary artery disease, cerebrovascular accident, peripheral vascular disease, cancer, and chronic obstructive pulmonary disease). In-hospital COVID-19 cases were identified from Medicare inpatient, outpatient, skilled nursing facility, home health agency, hospice, and physician/supplier claims in 2020 \citep{wu2022impact}. An in-hospital COVID-19 diagnosis was confirmed if any of these claim types associated with the inpatient stay contained either of the two primary diagnosis ICD-10 codes: B97.29 (other coronavirus as the cause of diseases classified elsewhere, since February 20, 2020) or U07.1 (COVID-19, from April 1, 2020 onward). Pediatric patients aged under 18 were excluded given their distinct characteristics compared to adult patients. After these exclusions, the data set included 683,328 discharges for 277,397 beneficiaries associated with 5,852 dialysis facilities.

\subsection{COVID-19 on unplanned hospital readmissions}
\begin{figure}[hbtp]
    \centering
    \includegraphics[width=0.65\textwidth]{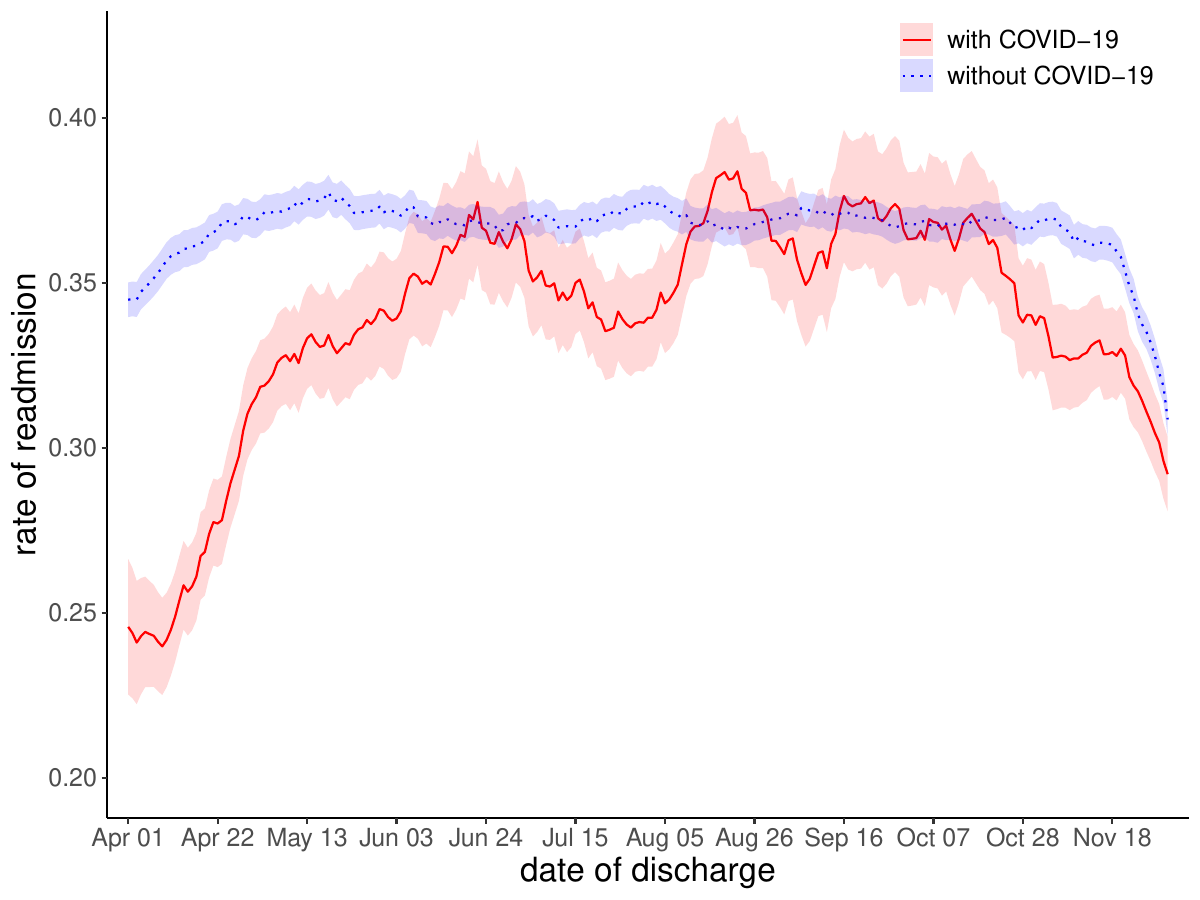}
    \caption{Rolling average (a window of 21 days) rates of 30-day all-cause unplanned hospital readmission among discharges with and without in-hospital COVID-19 between April 1 and December 1, 2020.}
    \label{fig:var_RR_covid}
\end{figure}

We first examined the impact of COVID-19 on 30-day unplanned readmissions, and the outcomes are presented in \Cref{fig:var_RR_covid}. During the period from early April to early June, discharges with in-hospital COVID-19 were linked to a significantly larger surge in the readmission rate compared to discharges without COVID-19. Following a brief decline, the readmission rate among COVID-19-related discharges rebounded, tracking closely with the rate among non-COVID-19 discharges, until approximately mid-October. Subsequently, the readmission rate for COVID-19-related discharges experienced a sharper decrease than the rate for discharges without COVID-19. This observation indicates that the effect of COVID-19 varied based on the date of discharge, warranting the need to employ the GPLM to effectively account for the dynamic effect trajectory.

Next, we fit both the GPLM and GLM to the Medicare claims data for dialysis patients. To ensure numerical stability, we further excluded dialysis facilities with fewer than 15 discharges, resulting in a final data set comprising 594,927 discharges for 242,608 beneficiaries across 3,016 dialysis facilities. In the GPLM, the FNN had 53 nodes in the input layer, 32 and 16 nodes in the following two hidden layers, respectively, and a single node in the output layer. As before, the activation functions were ReLU, ReLU, and identity, respectively. Counts and proportions for all levels of each risk factor, along with the corresponding odds ratios and 95\% confidence intervals from the GLM, are detailed in Table 3 of the Supplementary Material. Notably, COVID-19 exhibited an odds ratio of 0.804, indicating an inverse association between COVID-19 and 30-day readmission.

\subsection{Profiling kidney dialysis facilities}
\begin{figure}[hbtp]
    \centering
    \includegraphics[width=0.9\textwidth]{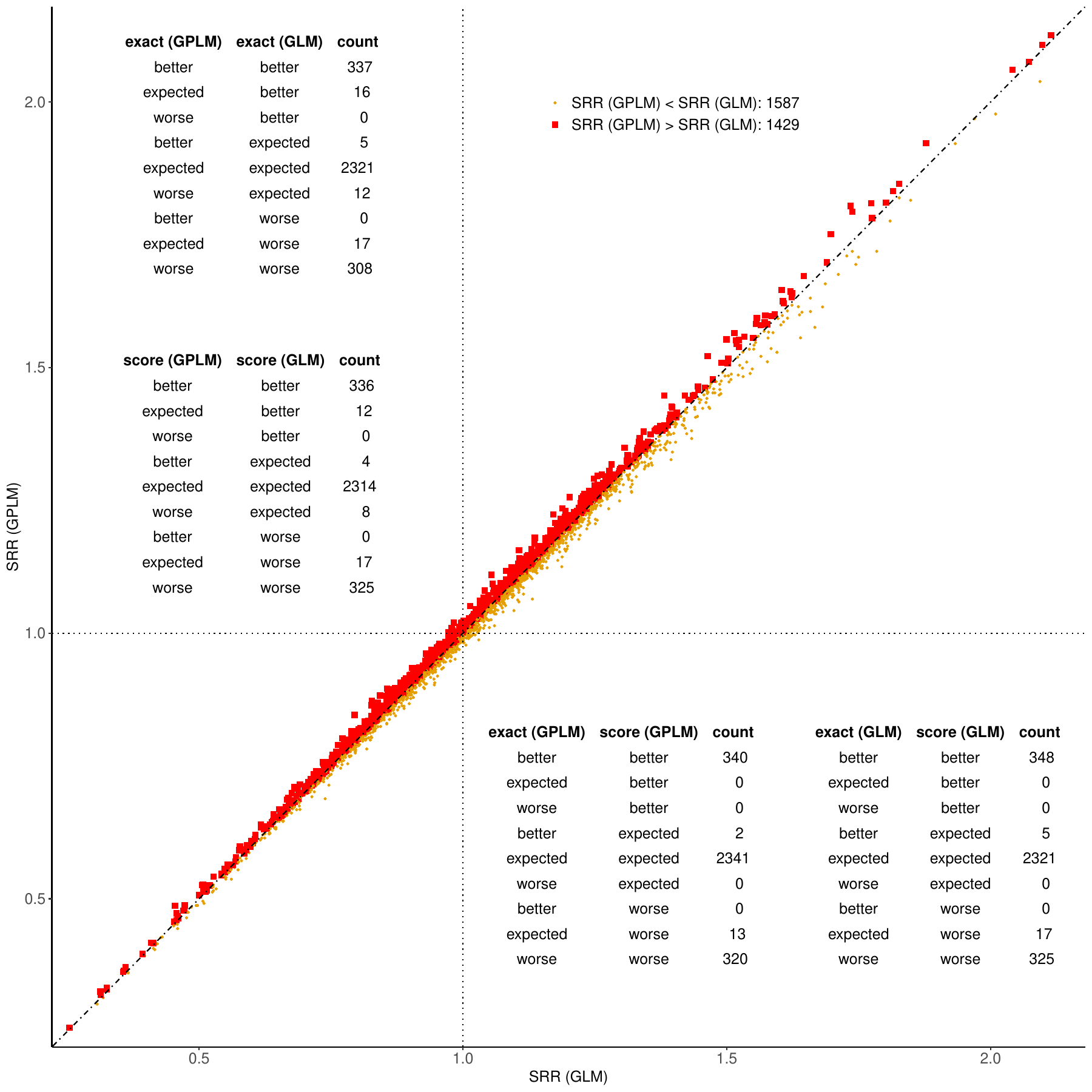}
    \caption{A scatter plot of standardized readmission ratios (SRRs) from the generalized partially linear model (GPLM) and generalized linear model (GLM). SRRs from the GPLM were calculated following \eqref{eq:measure}, while SRRs from the GLM were calculated following \citet{he2013evaluating} and \citet{wu2022improving, wu2023test}. Cross-tabulations of facility flagging based on the exact and score tests against the null hypothesis $H_{0,i}: \gamma_i = \mathrm{median}(\bgamma)$ from the GPLM and GLM are presented in the plot.}
    \label{fig:SRR_scatter_without_overdispersion}
\end{figure}

To profile Medicare-certified dialysis facilities, we computed the indirectly standardized ratio of unplanned readmission, also known as the standardized readmission ratio \citep[SRR,][]{he2013evaluating}, for each facility based on the GPLM and GLM, respectively. Shown in Figure 2 of the Supplementary Material, a facility with a high rate of readmission tends to have a high SRR. Additionally, we performed exact and score tests regarding the null hypothesis $H_{0,i}: \gamma_i = \mathrm{median}(\bgamma)$ for the two models. A facility is flagged as performing better (or worse) than expected if the null hypothesis is rejected and the SRR is less (or greater) than one. If the null is not rejected, the facility is flagged as performing as expected. The comparison between the two models is visualized in a scatter plot that displays GPLM-based and GLM-based SRRs (\Cref{fig:SRR_scatter_without_overdispersion}), accompanied by cross-tabulations of facility flagging.

Overall, the SRRs yielded by the two models are similar. GPLM-based SRRs ranged from 0.257 to 2.126, with a median and mean of 0.999 and 1.014, respectively; GLM-based SRRs ranged from 0.254 to 2.114, with a median and mean of 1.000 and 1.015, respectively. Among the 3,016 facilities, 1,587 (or 1,429) had GPLM-based SRRs greater (or less) than the GLM-based SRRs.

Few flagging discrepancies were observed between exact and score tests within the same model, or between corresponding tests from the GPLM and GLM. Specifically, GLM-based exact tests flagged 325 facilities (10.78\%) as performing worse than expected and 353 facilities (11.70\%) as performing better than expected. On the other hand, GPLM-based exact tests flagged 320 facilities (10.61\%) as worse and 342 facilities (11.34\%) as better than expected, with the proportion of outliers slightly lower than that from GPLM-based exact tests. A similar pattern emerges for score tests, suggesting that the GPLM offers a more flexible risk adjustment compared to the GLM.

\begin{figure}[hbtp]
    \centering
\includegraphics[width=\textwidth]{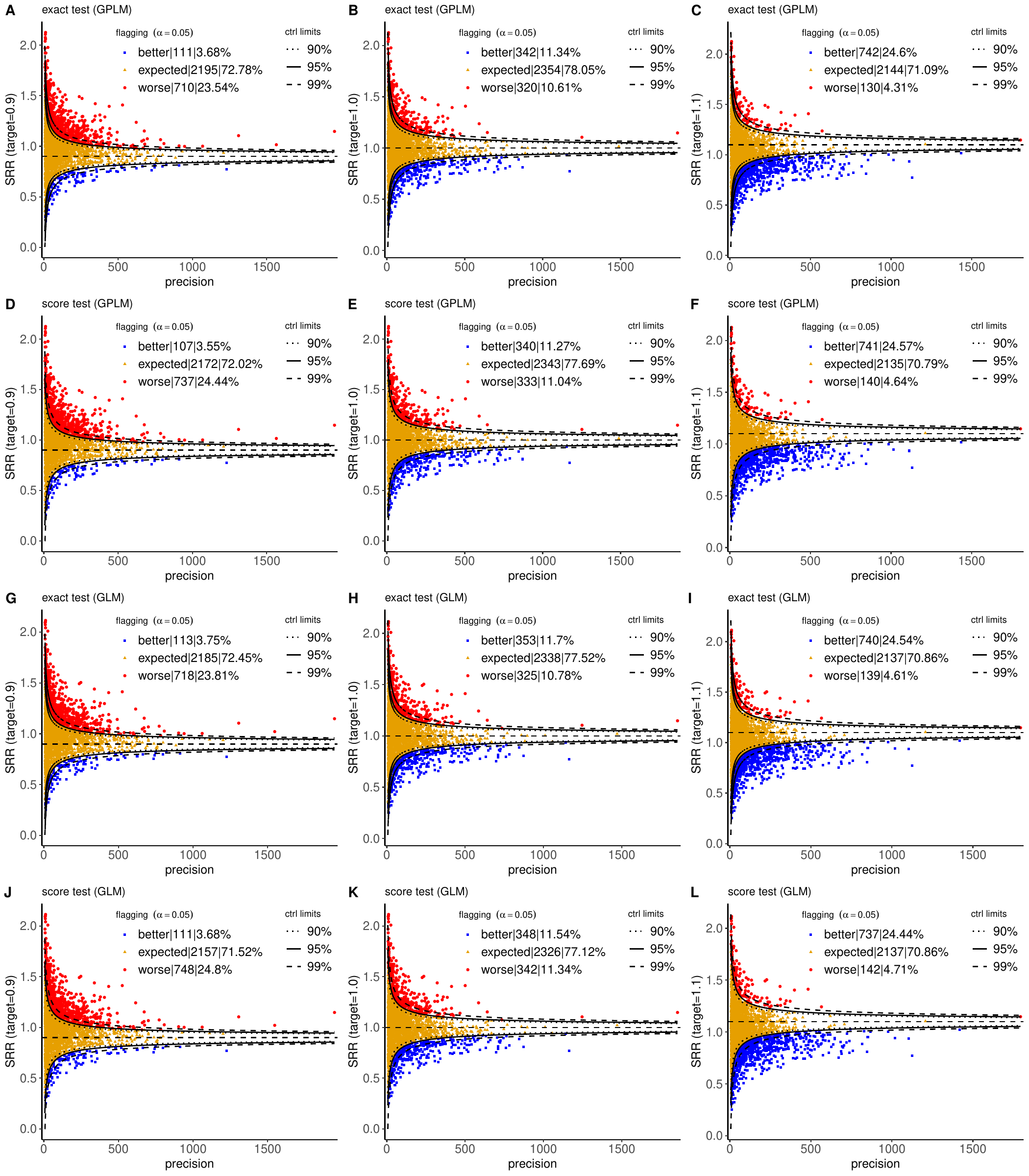}
    \caption{Funnel plots based on exact and score tests for the generalized partially linear model (GPLM) and the generalized linear model (GLM). Flagging corresponds to control limits for a $p$-value of 0.05.}
    \label{fig:funnels_without_overdispersion}
\end{figure}

The funnel plots depicted in \Cref{fig:funnels_without_overdispersion} display profiling results based on exact and score tests for both the GPLM and GLM. The columns represent increasing targets, while the $p$-values for control limits vary. As corroborated by previous findings \citep{silber2010hospital, horwitz2015association}, facilities with higher precision (also referred to as effective provider size) tend to exhibit a shorter span between control limits, which in turn increases the likelihood of them being flagged as outliers. Additionally, raising the target value corresponds to an increase in the number of better-performing facilities and a decrease in the number of worse-performing ones. Consistent with early observations, the proportion of outliers identified by either test within the GPLM is slightly lower than that within the GLM. Minimal disparities in flagging are noted between the exact and score tests.

\subsection{Accounting for unmeasured confounding}

A major issue of the profiling in \Cref{fig:SRR_scatter_without_overdispersion,fig:funnels_without_overdispersion} is that the proportion of facilities with an unusual performance identified by both tests is always greater than 20\%, much higher than what is typically anticipated in practice. In previous work, this problem has been recognized as unmeasured confounding that often leads to excess variation of provider-specific standardized quality measures \citep{spiegelhalter2005funnel, he2013evaluating, kalbfleisch2013monitoring, xia2022accounting, wu2022improving}. For example, due to the unavailability of secondary diagnosis information in the USRDS standard analysis files, many prevalent comorbidities were not accounted for in the risk adjustment of the GPLM and GLM. In addition, socioeconomic factors known to affect the risk of readmission 
, such as housing, income level, and educational attainment, are generally absent in Medicare claims.

Due to incomplete risk adjustment, a substantial portion of the variation in the outcome falls outside the control of providers, and addressing this overdispersion is imperative in provider profiling \citep{jones2011identification, kalbfleisch2018discussion}. Various statistical methods have been proposed to mitigate the impact of overdispersion.
In what follows, we employ the individualized empirical null (indivEN) method to determine the control limits of funnel plots due to its robustness in addressing overdispersion by linking the effective provider size with the marginal variance of standardized scores.

Rather than model \eqref{eq:parlincom}, here we assume that $\omega^*_{ij} = \gamma_i + \varphi_i + g^*(\bZ_{ij})$, where $\varphi_i \sim \cN(0, \sigma^2_\varphi)$ is a facility-specific random effect that could potentially account for overdispersion.
After obtaining the estimated lower-tail probability $G_i(O_i \mid \bZ_i; \mathrm{median}(\hat\bgamma), \hat\bw, \hat\bb, \tau)$ for facility $i$, we can express the Z-score as $X_i = \Phi^{-1}\{G_i(O_i \mid \bZ_i; \mathrm{median}(\hat\bgamma), \hat\bw, \hat\bb)\}$, with $\Phi$ denoting the distribution function of the standard Gaussian distribution. Further, assume that $X_i$ follows a mixture distribution, i.e., $X_i \sim \kappa_0 \cN(0, \chi_i^2) + (1-\kappa_0) \cN(\zeta_i, \iota_i^2)$, where $\kappa_0 \in (0, 1)$ denotes the proportion of facilities performing as expected. The variance $\chi_i^2$ can be well approximated by $\chi_i^2 \approx 1 + \sigma^2_\varphi V_i(\tau)$ with $V_i(\tau) = \tau \sum_{j=1}^{n_i} \dot{h}\{\mathrm{median}(\hat\bgamma) + g(\bZ_{ij}; \hat\bw, \hat\bb)\} [1 - \tau\dot{h}\{\mathrm{median}(\hat\bgamma) + g(\bZ_{ij}; \hat\bw, \hat\bb)\}]$ for Bernoulli outcomes. The parameters $\zeta_i$ and $\iota_i^2$ are treated as nuisance parameters. The estimation of the proportion $\kappa_0$ and variance $\sigma^2_\varphi$ can be achieved using the maximum likelihood approach \citep{efron2007size, xia2022accounting, hartman2023composite}. As a result, the control limits for $p$-value $\alpha$ and target $\tau$ are given by
\[
\left[\tau - \frac{\hat\chi_i z_{1-\alpha/2} \sqrt{V_i(\tau)}}{\sum_{j=1}^{n_i} \dot{h}\{\mathrm{median}(\hat\bgamma) + g(\bZ_{ij}; \hat\bw, \hat\bb)\}}, \tau + \frac{\hat\chi_i z_{1-\alpha/2} \sqrt{V_i(\tau)}}{\sum_{j=1}^{n_i} \dot{h}\{\mathrm{median}(\hat\bgamma) + g(\bZ_{ij}; \hat\bw, \hat\bb)\}}\right],
\]
where $z_{1-\alpha/2}$ denotes the $100\cdot(1-\alpha/2)$th percentile of the standard Gaussian distribution.

\begin{figure}[hbtp]
    \centering
\includegraphics[width=0.9\textwidth]{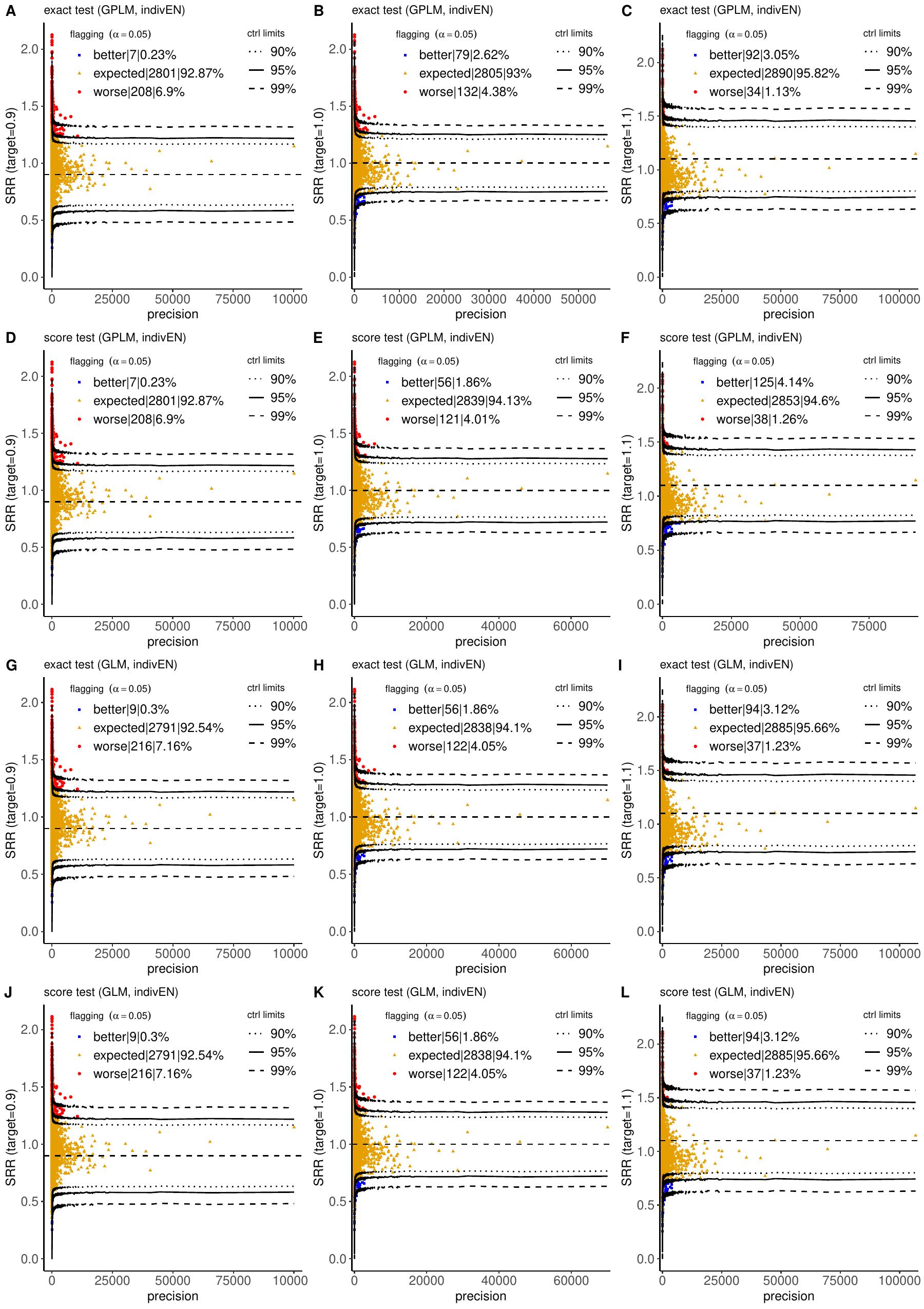}
    \caption{Funnel plots with overdispersion-adjusted control limits based on exact and score tests for the generalized partially linear model (GPLM) and the generalized linear model (GLM). Flagging corresponds to control limits for a $p$-value of 0.05. Control limits were determined following the individualized empirical null (indivEN) approach \citep{hartman2023composite}.}
    \label{fig:funnels_with_overdispersion}
\end{figure}

Just as shown in \Cref{fig:funnels_without_overdispersion}, \Cref{fig:funnels_with_overdispersion} illustrates funnel plots that incorporate overdispersion-adjusted control limits, based on exact and score tests for both the GPLM and GLM. Upon implementing the indivEN approach, the proportion of better- and worse-performing facilities across various scenarios was significantly diminished. For the GPLM-based exact test, 79 facilities (2.62\%) were classified as better-performing, and 132 (4.38\%) were identified as worse-performing. Similarly, the GPLM-based score test flagged 56 (1.86\%) facilities as better-performing and 121 (4.01\%) as worse-performing. The GLM-based tests indicated 56 (1.86\%) facilities as better-performing and 122 (4.05\%) as worse-performing. As previously observed, an increase in the target corresponds to a heightened proportion of better-performing facilities and a decreased proportion of worse-performing ones.

\section{Discussion}
\label{s:dis}

The significant undertaking of care quality assessment for providers has incentivized an enormous number of studies on advancing the methodology of providing profiling. However, no prior work has delved into addressing the potentially nonlinear associations between risk factors and outcomes for risk adjustment. While alternative flexible methods, such as the generalized additive and varying coefficient models, have found extensive use in capturing complex individual effect patterns in diverse applications, they are arguably not the most efficient solution within a profiling context. On the one hand, the numerical approximation tools (like kernel and spline functions) that underlie these models require a substantial amount of data to accurately depict multi-dimensional effect trajectories. This necessity presents formidable computational challenges, even when handling moderately large sample sizes (e.g., around half a million) and a dozen or so risk factors \citep{wu2022scalable, wu2022understanding}. On the other hand, while it is indeed crucial to accurately quantify and interpret individual effects of risk factors in statistical modeling aiming to identify a modifiable risk factor or examine the causal impact of an intervention, effect quantification or interpretation is typically not the primary focus in profiling.
This is especially the case when the overarching goal is to determine whether a provider's performance meets certain benchmarks. The goal would rather warrant a holistic approach that efficiently accounts for all possible interactions between risk factors as well as their potentially varying main effects. Consequently, this article represents a timely contribution that enriches profiling methods through the integration of efficient and powerful deep-learning technology. Inspired by \citet{mandel2023neural} and drawing on \citet{wu2022improving, wu2023test}, the proposed GPLM, along with the SSAMSGrad algorithm, the exact test, and funnel plots, collectively forms a streamlined toolkit that could potentially enhance the paradigm of profiling practice.

The development of the deep learning framework was motivated by a CMS dialysis facility profiling request in response to the COVID-19 pandemic, even though a different data source was utilized \citep{wu2022impact, wu2022understanding}. The overarching goal was to inform the CMS Dialysis Facility Care Compare program about the influence of the pandemic on the risk adjustment of SRR for dialysis facilities--a key CMS ESRD measure previously endorsed by the National Quality Forum \citep{kecc2017srr}. The evolving impact of COVID-19 on unplanned readmissions since the pandemic's onset, as demonstrated in \Cref{fig:var_RR_covid}, underscores the adoption of the flexible GPLM over the conventionally used GLM to enhance the characterization of the dynamic effect of COVID-19. The resulting improved risk adjustment likely contributes to the more conservative evaluations by the GPLM and its associated exact test, particularly concerning the identification of underperforming providers. The cross-tabulations of \Cref{fig:SRR_scatter_without_overdispersion} indicate profiling discrepancies, suggesting that GPLM-based tests tend to label facilities classified by the GLM as better or worse performers as satisfactory ones, while the exact test tends to categorize facilities designated as worse performers by the score test as expected. Given the significant financial implications of being labeled an underperformer, it is always sensible to exercise prudence in profiling analysis. In this regard, the conservative deep learning approach is indeed desirable.

While the GPLM and exact test exhibit a conservative approach in detecting underperforming providers compared to the GLM and score test, respectively, the challenge of incomplete risk adjustment remains significant. This is evident from the unrealistically high number of outliers observed in \Cref{fig:SRR_scatter_without_overdispersion,fig:funnels_without_overdispersion}, as well as in other studies \citep[e.g.,][]{he2013evaluating, wu2022improving, wu2023test}. To address this issue, we have adopted the indivEN method \citep{hartman2023composite}, a recent extension of the EN method, which offers comparative advantages over existing techniques. Among the various methods designed to estimate the null distribution of standardized scores, the original EN method \citep{efron2004large, efron2007size} does not incorporate provider volume as a key factor in its estimation. The characteristic function approach \citep{jin2007estimating}, while theoretically sound, proves numerically unstable in our application. The smoothed EN method, which takes provider volume into account through stratification, cannot be used to create sufficient provider strata with stable estimation when the number of providers is moderate (e.g., around 200). Traditional additive and multiplicative methods for addressing overdispersion, as discussed in \citet{spiegelhalter2005funnel}, also lack the explicit inclusion of provider volume as a factor.

When formulating the quality metric for facility profiling concerning unplanned readmissions, we have exclusively employed the approach of indirect standardization. This method contrasts the actual number of observed readmissions with the expected count that would arise if all Medicare dialysis beneficiaries were treated at a nationally representative facility. In contrast, direct standardization takes a hypothetical stance, contrasting the observed beneficiary population with a hypothetical one that would result if all beneficiaries were treated at the facility under consideration. Despite being occasionally misunderstood by practitioners and stakeholders, indirect standardization has gained prominence in a variety of profiling initiatives. This is partly due to its numerical robustness when dealing with small providers \citep{lee2002standardization}, which are prevalent in our context. It is noteworthy that while indirect standardization may not fully account for the effect of case mix differences across providers, leading to potential biases in assessments \citep{george2017mortality}, excessive concern is not warranted when the case mix is relatively similar across providers and rare risk factors are not a significant factor. In such situations, the concern would likely shift to the potential impact of unmeasured confounding.

The application of profiling dialysis facilities for Medicare ESRD beneficiaries, albeit comprehensive, should be interpreted in light of certain limitations. We obtained Medicare inpatient claims from USRDS standard analysis files, which provide only the primary diagnosis code for each beneficiary, while all other diagnosis codes are unavailable. Consequently, it is likely that the prevalence of every comorbid condition considered in the analysis was underestimated. To minimize the impact of this limitation, we considered all available claim types when identifying in-hospital COVID-19 cases. Comparing \Cref{fig:var_RR_covid} with Figure 1c of \citet{wu2022understanding}, we observe that both figures show similar readmission rates among discharges without COVID-19. However, after June 2020, the readmission rate among COVID-19 discharges in \Cref{fig:var_RR_covid} was no longer consistently higher than that among discharges without COVID-19. It is important to note that, even without limited access to diagnosis codes, the results related to COVID-19 are susceptible to under-reporting and misdiagnosis in 2020, which may be attributed to inconsistent testing and quarantine policies across different states \citep{salerno2021covid}. As a last note, we only considered dialysis facilities with at least 15 discharges in the Medicare data application. This treatment was meant to circumvent the intractable issue of algorithm convergence that is also present in many other profiling studies.

Building upon the proposed deep learning framework, several promising opportunities exist that could significantly advance the statistical paradigm of profiling. Firstly, while it is often deemed default to consider 30-day readmission as a longitudinal outcome, modeling the time to readmission would make nuanced provider differentiation possible, even if two providers experience similar readmission burdens over the 30-day period. Moreover, in readmission-focused profiling, the impact of death, which immediately terminates the observation of any subsequent event including readmission, is often overlooked. This simplified treatment can lead to underestimated readmission rates and potentially biased assessments in favor of providers associated with a disproportionate rate of mortality \citep{wu2022analysis}. Our current endeavors, representing an inaugural contribution grounded in deep learning, set the stage for the incorporation of death as a competing risk. This effort would likely facilitate a fundamental shift from unidimensional readmission-focused assessments towards more comprehensive provider monitoring, accounting for both readmission and mortality \citep{haneuse2022measuring}. Secondly, we have leveraged the FNN as the chief workhorse for complex risk adjustment. As deep learning continues to find interesting applications in longitudinal and time-to-event contexts \citep{fan2021selective, zhou2022deep, zhong2022deep}, novel methodological insights into the profiling problem could emerge by exploring advanced neural network structures such as the generative adversarial network \citep{goodfellow2014gan}. Finally, the proposed profiling methods are contingent on certain distributional assumptions.
For instance, \eqref{eq:expfam} is similar to certain conditions underlying the GLM and generalized linear mixed model; the exact test for identifying outlying providers is also distribution-based. These assumptions may be further relaxed by pursuing an alternative semi-parametric approach to estimating the mean of the readmission outcome.

\section*{Acknowledgments}
The authors thankfully acknowledge support from the Alzheimer's Association (AARG-23-1077773), National Heart, Lung, and Blood Institute (R01HL168202), National Institute on Aging (K02AG076883), National Institute of Biomedical Imaging and Bioengineering (P41EB017183), National Institute of Diabetes and Digestive and Kidney Diseases (R01DK070869), and the Department of Population Health and Center for the Study of Asian American Health at the NYU Grossman School of Medicine (U54MD000538). In addition, the authors are grateful to Dr. Tao Xu at the University of Michigan for discussions about the application to Medicare ESRD beneficiaries on kidney dialysis.

\section*{Disclaimer}
The data reported here have been supplied by the United States Renal Data System \citep[USRDS,][]{usrds2022adr}. The interpretation and reporting of these data are the responsibility of the authors and in no way should be seen as an official policy or interpretation of the U.S. government.

\bibliography{ref.bib}

\section*{Appendices}

\begin{appendices}

\section{Partial Derivatives in the Calculation of the Gradient of the Loss Function}

Following the notation in Section 2.2, for $i = 1, \ldots{}, m$, $j = 1, \ldots{}, n_i$, and $k = 1, \ldots{}, p_{l+1}$, we have the following partial derivatives of the log-likelihood function $\ell(\btheta)$:
\[
\frac{\partial \ell}{\partial \gamma_i}
= \sum_{j=1}^{n_i} [Y_{ij} - \dot{h}\{\gamma_i + g(\bZ_{ij}; \bw, \bb)\}],
\]
\[
\frac{\partial \ell}{\partial a_{ij}^{(L+1)}}
= Y_{ij} - \dot{h}\{\gamma_i + g(\bZ_{ij}; \bw, \bb)\},
\]
\[
\frac{\partial a_{ijk}^{(l+1)}}{\partial \bw_k^{(l+1)}}
= \dot{g}_{l+1}(\bw_k^{(l+1)} \ba_{ij}^{(l)} + b_k^{(l+1)}) (\ba_{ij}^{(l)})^\top,
\]
\[
\frac{\partial \ba_{ij}^{(l+1)}}{\partial \bb^{(l+1)}}
= \diag\left(\dot{g}_{l+1}(\bw_1^{(l+1)} \ba_{ij}^{(l)} + b_1^{(l+1)}), \ldots{}, \dot{g}_{l+1}(\bw_{p_{l+1}}^{(l+1)} \ba_{ij}^{(l)} + b_{p_{l+1}}^{(l+1)})\right),
\]
\[
\frac{\partial \ba_{ij}^{(l+1)}}{\partial \ba_{ij}^{(l)}}
= \underbrace{\left[\dot{g}_{l+1}(\bw^{(l+1)} \ba_{ij}^{(l)} + \bb^{(l+1)}), \ldots, \dot{g}_{l+1}(\bw^{(l+1)} \ba_{ij}^{(l)} + \bb^{(l+1)})\right]}_{p_{l} \text{ columns}} \otimes \bw^{(l+1)},
\]
where $\dot{h}$ denotes the first-order derivative of the univariate function $h$, $\bw_k^{(l+1)}$ is the $k$th row of $\bw^{(l+1)}$, $\diag$ indicates the diagonalization of a vector, and $\otimes$ denotes the Kronecker product.

\section{The ``Dropout'' Procedure to Mitigate Model Overfitting}

\begin{figure}[hbtp]
\centering
\includegraphics[width=0.75\textwidth]{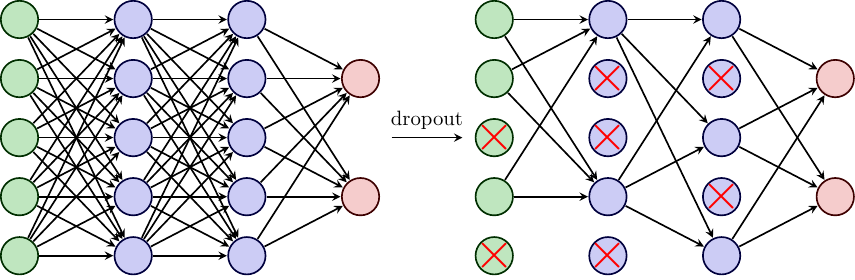}
\caption{The dropout procedure in training a feedforward neural network (FNN). A standard FNN on the left has two fully connected hidden layers, an input layer, and an output layer. Dropout is applied to all input and hidden layers, with the thinned FNN shown on the right. At each iteration, crossed nodes are dropped out.}
\label{fig:dropout}
\end{figure}

\section{Alternative Stochastic Optimization Algorithms}

\subsection{SSAdam: Stratified sampling adaptive moment estimation}

\begin{algorithm}[htbp]
\SetAlgoLined
\SetKwRepeat{Do}{do}{while}
\SetKw{KwFrom}{from}
initialize $s = 0$, $\psi \gg 10^8$, $\bgamma_{(0)} = \mathbf{0}$, $\bb_{(0)} = \mathbf{0}$, and $\br_{(0)} = \bv_{(0)} = \mathbf{0}$\;
\For{$l$ \KwFrom $0$ \KwTo $L+1$}{
$\bw^{(l+1)}_{(0)} \sim \mathrm{Uniform}(-\sqrt{6}/\sqrt{p_l + p_{l+1}}, \sqrt{6}/\sqrt{p_l + p_{l+1}})$\;
}
set $\delta \in (0.5, 1)$, $\eta = 10^{-3}$, $\xi \in (0, 1)$, $\beta_1 = 0.9$, $\beta_2 = 0.999$, $\epsilon = 10^{-8}$, and $u \in \mathbb{N}$\;
split $\cD$ into a training set $\cT$ and a validation set $\cV$ via stratified sampling with $|\cT| = \lfloor \delta |\cD| \rfloor$\;
\Do{$-|\cV|^{-1} \sum_{(i, j) \in \cV} \ell_{ij}(\btheta_{(s)}) > \psi$ {\rm across at most $u-1$ consecutive iterations}}{
$s \leftarrow s + 1$\;
$\bg_{(s)} = - |\cT_{(s)}|^{-1} \sum_{(i, j) \in \cT_{(s)}} \dot{\ell}_{ij}(\btheta_{(s-1)})$, where $\cT_{(s)}$ is a stratified random sample of $\cT$ with $|\cT_{(s)}| = \lfloor \xi |\cT| \rfloor$\;
$\br_{(s)} = \beta_1 \br_{(s-1)} + (1-\beta_1)\bg_{(s)}$\;
$\bv_{(s)} = \beta_2 \bv_{(s-1)} + (1-\beta_2) \bg_{(s)} \odot \bg_{(s)}$, where $\odot$ denotes the element-wise product\;
$\hat \br_{(s)} = (1 - \beta_1^s)^{-1}\br_{(s)}$\;
$\hat \bv_{(s)} = (1 - \beta_2^s)^{-1}\bv_{(s)}$\;
$\eta_{(s)} = \eta / \sqrt{s}$\;
$\btheta_{(s)} = \btheta_{(s-1)} - \eta_{(s)} \hat \br_{(s)} / (\sqrt{\hat \bv_{(s)}} + \epsilon)$, where square root and division are element-wise\;
$\psi \leftarrow \min\{\psi, -|\cV|^{-1} \sum_{(i, j) \in \cV} \ell_{ij}(\btheta_{(s)})\}$\;
}
\caption{SSAdam}
\label{alg:SSAdam}
\end{algorithm}

\subsection{SSRMSProp: Stratified sampling root mean square propagation}

\begin{algorithm}[htbp]
\SetAlgoLined
\SetKwRepeat{Do}{do}{while}
\SetKw{KwFrom}{from}
initialize $s = 0$, $\psi \gg 10^8$, $\bgamma_{(0)} = \mathbf{0}$, $\bb_{(0)} = \mathbf{0}$, and $\bv_{(0)} = \mathbf{0}$\;
\For{$l$ \KwFrom $0$ \KwTo $L+1$}{
$\bw^{(l+1)}_{(0)} \sim \mathrm{Uniform}(-\sqrt{6}/\sqrt{p_l + p_{l+1}}, \sqrt{6}/\sqrt{p_l + p_{l+1}})$\;
}
set $\delta \in (0.5, 1)$, $\eta = 10^{-3}$, $\xi \in (0, 1)$, $\beta = 0.9$, $\epsilon = 10^{-8}$, and $u \in \mathbb{N}$\;
split $\cD$ into a training set $\cT$ and a validation set $\cV$ via stratified sampling with $|\cT| = \lfloor \delta |\cD| \rfloor$\;
\Do{$-|\cV|^{-1} \sum_{(i, j) \in \cV} \ell_{ij}(\btheta_{(s)}) > \psi$ {\rm across at most $u-1$ consecutive iterations}}{
$s \leftarrow s + 1$\;
$\bg_{(s)} = - |\cT_{(s)}|^{-1} \sum_{(i, j) \in \cT_{(s)}} \dot{\ell}_{ij}(\btheta_{(s-1)})$, where $\cT_{(s)}$ is a stratified random sample of $\cT$ with $|\cT_{(s)}| = \lfloor \xi |\cT| \rfloor$\;
$\bv_{(s)} = \beta \bv_{(s-1)} + (1-\beta) \bg_{(s)} \odot \bg_{(s)}$, where $\odot$ denotes the element-wise product\;
$\eta_{(s)} = \eta / \sqrt{s}$\;
$\btheta_{(s)} = \btheta_{(s-1)} - \eta_{(s)} \bg_{(s)} / (\sqrt{\bv_{(s)}} + \epsilon)$, where square root and division are element-wise\;
$\psi \leftarrow \min\{\psi, -|\cV|^{-1} \sum_{(i, j) \in \cV} \ell_{ij}(\btheta_{(s)})\}$\;
}
\caption{SSRMSProp}
\label{alg:SSRMSProp}
\end{algorithm}

\subsection{SSSGD: Stratified sampling stochastic gradient descent}

\begin{algorithm}[htbp]
\SetAlgoLined
\SetKwRepeat{Do}{do}{while}
\SetKw{KwFrom}{from}
initialize $s = 0$, $\psi \gg 10^8$, $\bgamma_{(0)} = \mathbf{0}$, and $\bb_{(0)} = \mathbf{0}$\;
\For{$l$ \KwFrom $0$ \KwTo $L+1$}{
$\bw^{(l+1)}_{(0)} \sim \mathrm{Uniform}(-\sqrt{6}/\sqrt{p_l + p_{l+1}}, \sqrt{6}/\sqrt{p_l + p_{l+1}})$\;
}
set $\delta \in (0.5, 1)$, $\eta = 10^{-3}$, $\xi \in (0, 1)$, $\beta = 0.9$, $\epsilon = 10^{-8}$, and $u \in \mathbb{N}$\;
split $\cD$ into a training set $\cT$ and a validation set $\cV$ via stratified sampling with $|\cT| = \lfloor \delta |\cD| \rfloor$\;
\Do{$-|\cV|^{-1} \sum_{(i, j) \in \cV} \ell_{ij}(\btheta_{(s)}) > \psi$ {\rm across at most $u-1$ consecutive iterations}}{
$s \leftarrow s + 1$\;
$\bg_{(s)} = - |\cT_{(s)}|^{-1} \sum_{(i, j) \in \cT_{(s)}} \dot{\ell}_{ij}(\btheta_{(s-1)})$, where $\cT_{(s)}$ is a stratified random sample of $\cT$ with $|\cT_{(s)}| = \lfloor \xi |\cT| \rfloor$\;
$\eta_{(s)} = \eta / \sqrt{s}$\;
$\btheta_{(s)} = \btheta_{(s-1)} - \eta_{(s)} \bg_{(s)}$\;
$\psi \leftarrow \min\{\psi, -|\cV|^{-1} \sum_{(i, j) \in \cV} \ell_{ij}(\btheta_{(s)})\}$\;
}
\caption{SSSGD}
\label{alg:SSSGD}
\end{algorithm}

\section{Supplementary Tables for Simulations}

\begin{table}[htbp]
\caption{Performance of the SSAMSGrad, SSAdam, SSRMSProp, and SSSGD under generalized partially linear model (GPLM). For every predictive metric except time to convergence (runtime), the mean and standard deviation (in parentheses) of each metric are derived from 50 simulated data sets. Runtime is measured in seconds, and only the mean is provided. \label{tab:4stooptmethods}}
\centering
\begin{tabular}{ccccc}
\toprule
\toprule
& SSAMSGrad & SSAdam & SSRMSProp & SSSGD \\
\midrule
accuracy & 0.731 (0.010) & 0.731 (0.010) & 0.730 (0.010) & 0.721 (0.011) \\
sensitivity & 0.815 (0.017) & 0.813 (0.017) & 0.812 (0.019) & 0.807 (0.022) \\
specificity & 0.629 (0.025) & 0.630 (0.026) & 0.630 (0.027) & 0.615 (0.034) \\
precision & 0.729 (0.014) & 0.729 (0.013) & 0.729 (0.014) & 0.720 (0.015) \\
F1 & 0.769 (0.011) & 0.769 (0.011) & 0.768 (0.012) & 0.760 (0.012) \\
AUC & 0.800 (0.011) & 0.800 (0.011) & 0.799 (0.011) & 0.788 (0.012) \\
loss & 0.530 (0.010) & 0.530 (0.009) & 0.532 (0.010) & 0.548 (0.011) \\
runtime & 2.373 & 5.606 & 3.066 & 25.40 \\
\bottomrule
\bottomrule
\end{tabular}
\end{table}

\begin{table}[htbp]
\caption{Performance of the AMSGrad, Adam, and RMSProp under generalized partially linear model (GPLM). For each predictive metric, the mean and standard deviation (in parentheses) of each metric are derived from 500 simulated data sets. The average speedup is calculated as follows: (1) for each algorithm, the average time to convergence is measured across five model fits with the same simulated data; (2) speedups of the AMSGrad relative to the Adam and RMSProp are calculated respectively for each simulated data; (3) speedups are averaged across 50 simulated data sets. \label{tab:stooptmethodssim}}
\centering
\resizebox{\columnwidth}{!}{%
\begin{tabular}{cccccccccccccccc}
\toprule
\toprule
\multicolumn{8}{c}{Panel A: $g^*_1(\bZ_{ij})$} \\
\midrule
\multirow{3}{*}{$m$} & \multirow{3}{*}{metric} & \multicolumn{2}{c}{AMSGrad} & \multicolumn{2}{c}{Adam} & \multicolumn{2}{c}{RMSProp} \\
\cmidrule(r){3-4} \cmidrule(lr){5-6} \cmidrule(l){7-8}
& & $\nu=50$ & $\nu=100$ & $\nu=50$ & $\nu=100$ & $\nu=50$ & $\nu=100$ \\
\midrule
\multirow{7}{*}{100} 
 & accuracy & 0.745 (0.015) & 0.747 (0.011)  & 0.745 (0.015) & 0.747 (0.011) & 0.744 (0.015) & 0.746 (0.010) \\
 & sensitivity & 0.881 (0.021) & 0.886 (0.016) & 0.879 (0.019) & 0.884 (0.015) & 0.881 (0.019) & 0.885 (0.014)\\
 & specificity & 0.450 (0.043) & 0.445 (0.035) & 0.451 (0.042) & 0.449 (0.032) & 0.447 (0.044) & 0.444 (0.034) \\
 & precision & 0.777 (0.017) & 0.776 (0.013) & 0.778 (0.017) & 0.777 (0.013) & 0.776 (0.018) & 0.775 (0.012) \\
 & F1 & 0.825 (0.013) & 0.827 (0.010) & 0.825 (0.013) & 0.827 (0.010) & 0.825 (0.012) & 0.826 (0.009) \\
 & AUC & 0.774 (0.016) & 0.777 (0.011) & 0.774 (0.016) & 0.776 (0.011) & 0.773 (0.015) & 0.776 (0.012) \\
 & speedup & 1 & 1 & 2.630 & 2.541 & 1.093 & 1.051\\ 
\midrule
\multirow{7}{*}{300} 
 & accuracy & 0.746 (0.009) & 0.748 (0.007) & 0.747 (0.009) & 0.748 (0.006) & 0.745 (0.008) & 0.747 (0.006) \\
 & sensitivity & 0.886 (0.012) & 0.890 (0.009) & 0.885 (0.011) & 0.889 (0.009) & 0.885 (0.012) & 0.888 (0.009) \\
 & specificity & 0.443 (0.024) & 0.439 (0.020) & 0.447 (0.025) & 0.440 (0.020) & 0.441 (0.026) & 0.440 (0.020) \\
 & precision & 0.775 (0.010) & 0.775 (0.007) & 0.776 (0.010) & 0.775 (0.007) & 0.775 (0.009) & 0.775 (0.007) \\
 & F1 & 0.827 (0.008) & 0.829 (0.006) & 0.827 (0.008) & 0.828 (0.006) & 0.826 (0.007) & 0.828 (0.006) \\
 & AUC & 0.776 (0.009) & 0.778 (0.007) & 0.777 (0.010) & 0.777 (0.007) & 0.774 (0.009) & 0.777 (0.007) \\
 & speedup & 1 & 1 & 2.795 & 3.010 & 1.051 & 1.136 \\ 
\toprule
\multicolumn{8}{c}{Panel B: $g^*_2(\bZ_{ij})$} \\
\midrule
\multirow{3}{*}{$m$} & \multirow{3}{*}{metric} & \multicolumn{2}{c}{AMSGrad} & \multicolumn{2}{c}{Adam} & \multicolumn{2}{c}{RMSProp}  \\
\cmidrule(r){3-4} \cmidrule(lr){5-6} \cmidrule(lr){7-8}
& & $\nu=50$ & $\nu=100$ & $\nu=50$ & $\nu=100$ & $\nu=50$ & $\nu=100$ \\
\midrule
\multirow{7}{*}{100} 
 & accuracy & 0.729 (0.015) & 0.731 (0.010) & 0.728 (0.014) & 0.731 (0.010) & 0.728 (0.015) & 0.731 (0.010)  \\
 & sensitivity & 0.809 (0.024) & 0.815 (0.017) & 0.805 (0.026) & 0.810 (0.018) & 0.807 (0.025) & 0.814 (0.019) \\
 & specificity & 0.630 (0.034) & 0.627 (0.026) & 0.631 (0.032) & 0.633 (0.026) & 0.630 (0.033) & 0.628 (0.025) \\
 & precision & 0.728 (0.020) & 0.729 (0.014) & 0.730 (0.019) & 0.730 (0.014) & 0.729 (0.019) & 0.729 (0.014) \\
 & F1 & 0.766 (0.016) & 0.769 (0.011) & 0.766 (0.015) & 0.768 (0.012) & 0.766 (0.015)  & 0.769 (0.012) \\
 & AUC & 0.797 (0.015) & 0.800 (0.011) & 0.797 (0.014) & 0.800 (0.011) & 0.796 (0.015) & 0.801 (0.010) \\
& speedup & 1 & 1 & 3.096 & 2.656 & 1.115 & 0.982 \\ 
\midrule
\multirow{7}{*}{300} 
 & accuracy & 0.730 (0.008) & 0.732 (0.006) & 0.730 (0.008) & 0.732 (0.006)  & 0.730 (0.008) & 0.732 (0.006) \\
 & sensitivity & 0.813 (0.014) & 0.818 (0.010) & 0.812 (0.014) & 0.817 (0.011) & 0.812 (0.014) & 0.818 (0.010) \\
 & specificity & 0.628 (0.019) & 0.627 (0.014) & 0.628 (0.018) & 0.628 (0.015) & 0.627 (0.019) & 0.627 (0.015) \\
 & precision & 0.728 (0.011) & 0.729 (0.008) & 0.729 (0.011) & 0.730 (0.008) & 0.729 (0.011) & 0.729 (0.008) \\
 & F1 & 0.768 (0.009) & 0.771 (0.006) & 0.768 (0.009) & 0.771 (0.007) & 0.768 (0.009) & 0.771 (0.006)\\
 & AUC & 0.800 (0.008) & 0.802 (0.006) & 0.799 (0.009) & 0.801 (0.007) & 0.798 (0.008) & 0.802 (0.006) \\
 & speedup & 1 & 1 & 2.877 & 2.755 & 1.042 & 0.707 \\ 
\bottomrule
\bottomrule
\end{tabular}
}
\end{table}

\section{ICD-10 Codes to Identify Planned Hospital Readmissions}

Z44001, Z44002, Z44009, Z44011, Z44012, Z44019, Z44021, Z44022, Z44029, Z44101, Z44102, Z44109, 
Z44111, Z44112, Z44119, Z44121, Z44122, Z44129, Z4430, Z4431, Z4432, Z448, Z449, Z451, Z4531, Z45320, 
Z45321, Z45328, Z4541, Z4542, Z4549, Z45811, Z45812, Z45819, Z4682, Z4689, Z469, Z510, Z5111, and Z5112.

\section{Supplementary Table for the Application to Medicare Beneficiaries on Kidney Dialysis in 2020}

\begin{longtable}[htbp]{rrrr}
\caption{Summary of fitting the generalized linear model to the Medicare inpatient claims data for beneficiaries with end-stage renal disease (ESRD) undergoing kidney dialysis in 2020. Levels of covariates in parentheses represent reference groups. OR stands for the odds ratio. NHPI stands for Native Hawaiian \& Pacific Islander. COPD stands for chronic obstructive pulmonary disease. CAPD stands for continuous ambulatory peritoneal dialysis. CCPD stands for continuous cycling peritoneal dialysis.} \label{tab:summary_covar} \\
\toprule
\toprule
\multicolumn{1}{l}{Covariates} & Count & Proportion & OR (lower limit, upper limit) \\
\midrule
\endfirsthead
\multicolumn{4}{c}%
{{\tablename\ \thetable{} -- continued}} \\
\midrule
\multicolumn{1}{l}{Covariates} & Count & Proportion & OR (lower limit, upper limit) \\
\midrule
\endhead
\multicolumn{4}{r}{{continued on next page}} \\
\midrule
\endfoot
\endlastfoot
\multicolumn{1}{l}{Race (White)} & \multicolumn{3}{c}{} \\
Black & 192008 & 32.31\% &  0.998 (0.983, 1.013) \\
 Asian & 18671 & 3.14\% & 0.825 (0.796, 0.855) \\
NHPI & 5653 & 0.95\% & 0.863 (0.810, 0.919) \\ 
Other & 10030 & 1.69\% & 0.947 (0.901, 0.995)\\
\multicolumn{1}{l}{Female} & 267952 & 45.1\% & 1.033 (1.022, 1.045) \\ 
\multicolumn{1}{l}{Hispanic} & 83462 & 14.0\% & 0.911 (0.894, 0.930) \\ 
  \multicolumn{1}{l}{Age in years (50--64)} & \multicolumn{3}{c}{} \\
 18--34 & 50174 & 8.44\% & 1.293 (1.265, 1.322) \\ 
 35--49 & 109153 & 18.4\% & 1.062 (1.045, 1.080) \\ 
 65--74 & 139412 & 23.5\% & 0.905 (0.890, 0.919) \\ 
 $\ge$74 & 92233 & 15.5\% & 0.831 (0.816, 0.847) \\ 
  \multicolumn{1}{l}{BMI ($\ge$30)} & \multicolumn{3}{c}{} \\
 $\le$18.5 & 15281 & 2.57\% & 1.137 (1.097, 1.177) \\
 18.5--25 & 157388 & 26.5\% & 1.076 (1.061, 1.091) \\
 25--30 & 163124 & 27.5\% & 1.052 (1.037, 1.066) \\
\multicolumn{1}{l}{Drug use} & 10084 & 1.70\% & 1.346 (1.290, 1.405) \\
\multicolumn{1}{l}{Alcohol use} & 8908 & 1.50\% & 1.130 (1.080, 1.182) \\
\multicolumn{1}{l}{Tobacco use} & 42108 & 7.09\% & 1.107 (1.083, 1.132) \\
\multicolumn{1}{l}{Heart failure} & 161161 & 27.1\% & 1.052 (1.039, 1.066)\\ 
\multicolumn{1}{l}{Coronary artery disease} & 71848 & 12.1\% & 1.012 (0.994, 1.031) \\ 
\multicolumn{1}{l}{Cerebrovascular accident} & 49889 & 8.40\% & 0.993 (0.974, 1.014) \\ 
\multicolumn{1}{l}{Peripheral vascular disease} & 54094 & 9.10\% & 1.022 (1.002, 1.043) \\ 
\multicolumn{1}{l}{Cancer} & 34296 & 5.77\% & 1.012 (0.988, 1.036) \\ 
\multicolumn{1}{l}{COPD} & 49644 & 8.35\% & 1.098 (1.075, 1.120) \\ 
\multicolumn{1}{l}{Employment (Retired)} & \multicolumn{3}{c}{} \\ 
\quad Unemployed & 161998 & 27.3\% & 1.025 (1.010, 1.039)\\ 
\quad Employed & 72199 & 12.2\% & 0.854 (0.838, 0.871)\\ 
\quad Other employment & 31771 & 5.35\% & 0.853 (0.831, 0.876) \\
\multicolumn{1}{l}{Length of hospital stay ($\le$3)} & \multicolumn{3}{c}{} \\ 
\quad 3--5 & 206533 & 34.8\% & 1.133 (1.115, 1.150) \\ 
\quad 6--9 & 129998 & 21.9\% & 1.361 (1.338, 1.385) \\
\quad $\ge$9 & 128933 & 21.7\% & 1.682 (1.652, 1.712) \\ 
\multicolumn{1}{l}{ESRD vintage ($\ge$5 years)} & \multicolumn{3}{c}{} \\ 
\quad $\le$90 days & 74297 & 12.5\% & 1.161 (1.138, 1.185) \\ 
\quad 90--182 days & 19558 & 3.29\% & 1.041 (1.008, 1.075) \\ 
\quad $182$ days to $1$ year & 38434 & 6.47\% & 1.047 (1.022, 1.073) \\ 
\quad 1--2 years & 72250 & 12.2\% & 1.072 (1.052, 1.093) \\ 
\quad 2--3 years & 65060 & 10.9\% & 1.081 (1.060, 1.102) \\
\quad 3--5 years & 107173 & 18.0\% & 1.081 (1.064, 1.099) \\
\multicolumn{1}{l}{Functionally limited} & 92593 & 15.6\% & 1.062 (1.045, 1.079) \\ 
\multicolumn{1}{l}{Cause of ESRD (Diabetes)} & \multicolumn{3}{c}{} \\ 
\quad Primary glomerulonephritis & 52009 & 8.75\% & 0.926 (0.906, 0.946) \\
\quad Hypertension & 167377 & 28.2\% & 0.951 (0.938, 0.964) \\ 
\quad Other & 86638 & 14.6\% & 0.962 (0.946, 0.979) \\ 
\multicolumn{1}{l}{Dialysis mode (Hemodialysis)} & \multicolumn{3}{c}{} \\ 
\quad CAPD & 22549 & 3.79\% & 0.952 (0.924, 0.980) \\ 
\quad CCPD & 27049 & 4.55\% & 0.948 (0.923, 0.975) \\ 
\quad Other & 5706 & 0.96\% & 0.806 (0.759, 0.857) \\ 
\multicolumn{1}{l}{In-hospital COVID-19} & 42802 & 7.20\% & 0.804 (0.786, 0.823) \\ 
\multicolumn{1}{l}{Month (January)} & \multicolumn{3}{c}{} \\
\quad February & 55406 & 9.32\% & 0.967 (0.944, 0.991) \\ 
\quad March & 52676 & 8.86\% & 0.836 (0.816, 0.857) \\ 
\quad April & 40976 & 6.90\% & 0.942 (0.917, 0.967) \\ 
\quad May & 45394 & 7.64\% & 1.027 (1.001, 1.053) \\ 
\quad June & 47735 & 8.03\% & 1.020 (0.994, 1.046) \\ 
\quad July & 49867 & 8.39\% & 1.028 (1.003, 1.054) \\ 
\quad August & 48358 & 8.14\% & 1.023 (0.998, 1.049) \\ 
\quad September & 47823 & 8.05\% & 1.040 (1.014, 1.067) \\ 
\quad October & 49456 & 8.32\% & 1.029 (1.003, 1.055) \\ 
\quad November & 45235 & 7.61\% & 0.984 (0.959, 1.010) \\ 
\quad December & 49650 & 8.36\% & 0.351 (0.341, 0.361) \\
\bottomrule
\end{longtable}

\section{Supplementary Figure for the Application to Medi-
care Beneficiaries on Kidney Dialysis in 2020}

\begin{figure}[hbtp]
    \centering
    \includegraphics[width=0.8\textwidth]{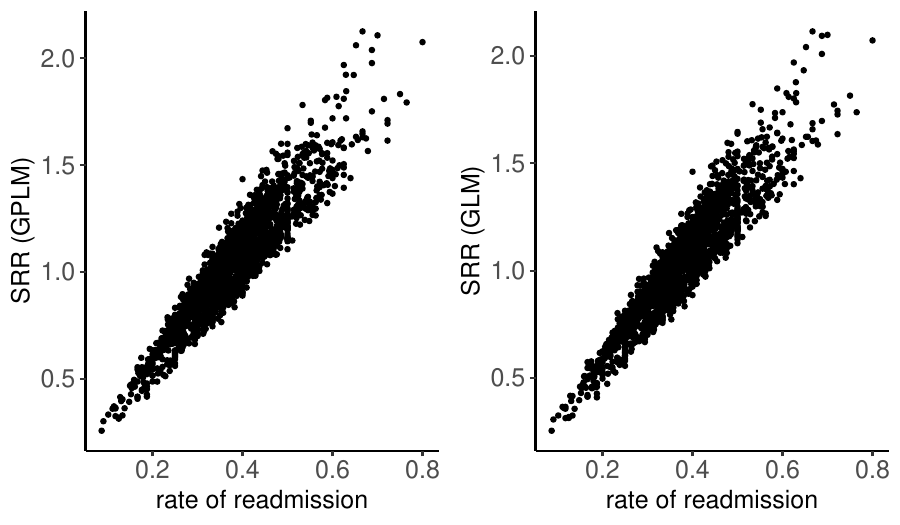}
    \caption{Scatter plots of the facility-specific standardized readmission ratio (SRR) derived from the generalized partially linear model (GPLM) and the generalized linear model (GLM) versus the rate of readmission.}
    \label{fig:SRRvsRRscattern}
\end{figure}

\end{appendices}

\end{document}